\documentclass[12pt]{iopart}

\usepackage{iopams}  
\usepackage{pgf}
\usepackage{pdfpages}
\usepackage{graphicx}
\usepackage{float}
\usepackage[caption=false]{subfig}
\DeclareMathAlphabet{\mathdutchcal}{U}{dutchcal}{m}{n} 
\SetMathAlphabet{\mathdutchcal}{bold}{U}{dutchcal}{b}{n}
\DeclareMathAlphabet{\mathdutchbcal}{U}{dutchcal}{b}{n} 

\begin{document}

\title[EVS and Arcsine Laws of Brownian Motion with a Permeable Barrier]{Extreme Value Statistics and Arcsine Laws of Brownian Motion in the Presence of a Permeable Barrier}

\author{Toby Kay$^1$ and Luca Giuggioli$^{1,2}$}

\address{$^1$ Department of Engineering Mathematics, University of Bristol, Bristol, BS8 1UB, United Kingdom}
\address{$^2$ Bristol Centre for Complexity Sciences, University of Bristol, Bristol, BS8 1UB, United Kingdom }
\ead{toby.kay@bristol.ac.uk}
\vspace{10pt}
\begin{indented}
\item[]\today
\end{indented}
\begin{abstract}
    The Arcsine laws of Brownian motion are a collection of results describing three different statistical quantities of one-dimensional Brownian motion: the time at which the process reaches its maximum position, the total time the process spends in the positive half-space and the time at which the process crosses the origin for the last time. Remarkably the cumulative probabilities of these three observables all follow the same distribution, the Arcsine distribution. But in real systems, space is often heterogeneous, and these laws are likely to hold no longer. In this paper we explore such a scenario and study how the presence of a spatial heterogeneity alters these Arcsine laws. Specifically we consider the case of a thin permeable barrier, which is often employed to represent diffusion impeding heterogeneities in physical and biological systems such as multilayer electrodes, electrical gap junctions, cell membranes and fragmentation in the landscape for dispersing animals. Using the Feynman-Kac formalism and path decomposition techniques we are able to find the exact time-dependence of the probability distribution of the three statistical quantities of interest. We show that a permeable barrier has a large impact on these distributions at short times, but this impact is less influential as time becomes long. In particular, the presence of a barrier means that the three distributions are no longer identical with symmetry about their means being broken. We also study a closely related statistical quantity, namely, the distribution of the maximum displacement of a Brownian particle and show that it deviates significantly from the usual half-Gaussian form.  
\end{abstract}

%
%
%
%
%

\section{Introduction}

Diffusion is ubiquitous, appearing as a transport mechanism in many physical, chemical and biological systems. Often these systems are littered with spatial heterogeneities which impede or hamper the diffusive motion. One of the most common forms of these heterogeneities are permeable barriers, such that diffusive particles either pass through or are reflected upon interaction with the barrier. These barriers appear at various scales inhibiting diffusive movement in many physical and biological contexts, from multilayer electrodes \cite{diard2005one,freger2005diffusion,ngameni2014derivation} and water transport in rock pores \cite{song2000determining} to drug delivery systems \cite{siegel1986laplace,pontrelli2007mass}. Permeable material can be found in many examples in cell biology whose function is to regulate the flux of biochemicals between spatial regions \cite{phillips2012physical}, such as the bilayer plasma membrane of eukaryotes \cite{kenkre2008molecular,kusumi2005paradigm,nikonenko2021ion} and the electrical gap junctions in neurons \cite{evans2002gap,connors2004electrical}. As well as being prominent in magnetic imaging techniques of water molecules diffusing through cellular compartments \cite{grebenkov2014exploring1,grebenkov2014exploring2}. Permeable substances also present themselves at larger scales such as heterogeneous landscapes, e.g. habitat type \cite{beyer2016you,kenkre2021theory} or the presence of roads and fences \cite{assis2019road}, affecting the dispersal of animal movement at ecological scales. All these examples make it apparent that it is necessary to build a mathematical understanding of how the diffusive movement statistics is affected by the presence of a permeable barrier. Here we do so by investigating how the extreme value statistics (EVS) and the closely related Arcsine laws of Brownian motion (BM) change with permeable barriers. 

\begin{figure}[htp]
    \centering
    \includegraphics[width=0.5\textwidth]{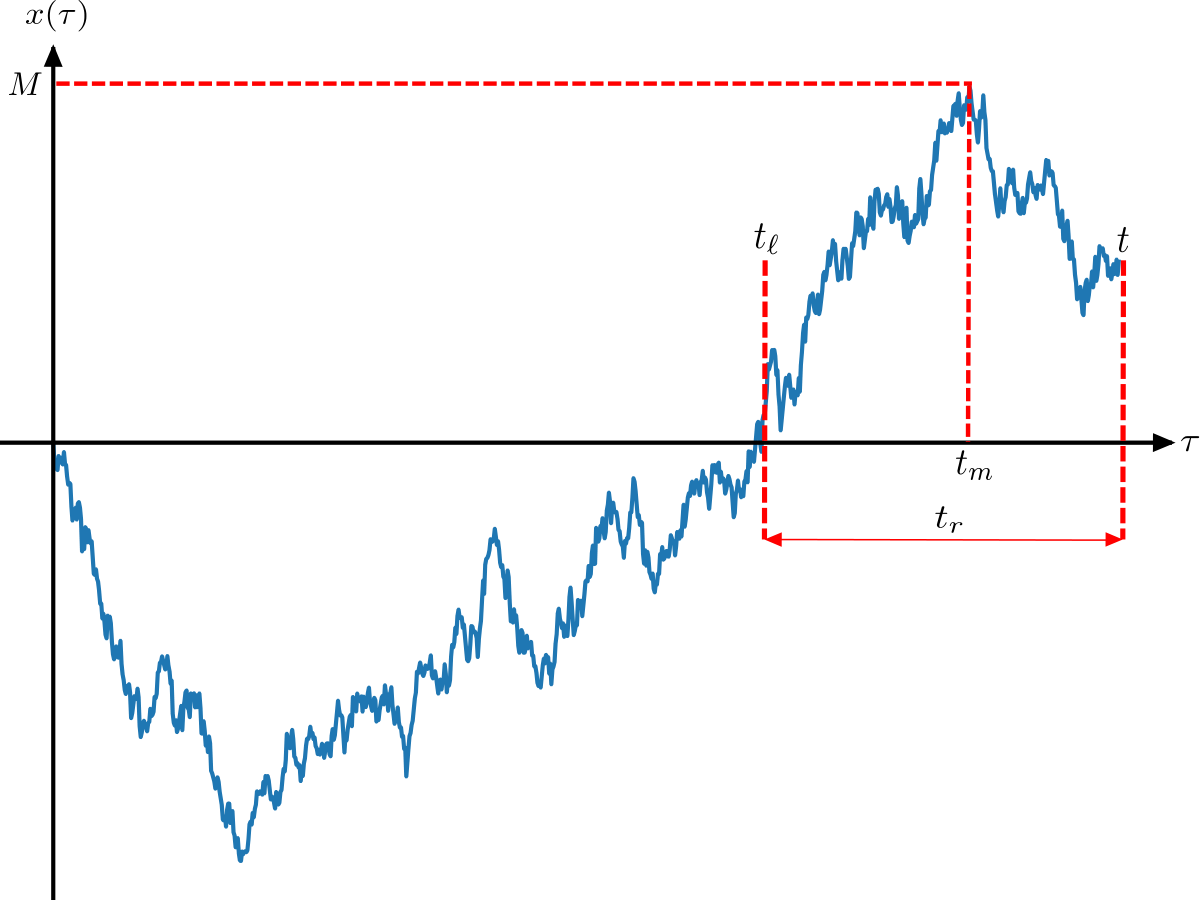}
    \caption{Illustration of the three observables, time of maximum position $x(t_m)=M$, residence time in the positive half space, $t_r$, time of last crossing of the origin, $t_\ell$, for a sample Brownian motion trajectory.}
    \label{fig:trajectory}
\end{figure}

The celebrated Arcsine laws of BM are a landmark result from L\'{e}vy \cite{levy1940certains} describing the probability distribution of three observables of a BM trajectory, $x(\tau)$, starting at the origin, over a time interval $\tau\in[0,t]$: (1.) $t_m(t)$, the time at which the trajectory reaches its maximum value, $x(t_m)=M$, (2.) $t_r(t)$, the total time the trajectory spends in the positive region, $x(\tau)>0$, (3.) $t_\ell(t)$, the time at which the trajectory crosses the origin for the last time. These quantities are displayed for a sample BM trajectory in figure \ref{fig:trajectory}. The remarkable feature of these quantities, $t_m(t)$, $t_r(t)$ and $t_\ell(t)$ is that they all have the same cumulative distribution function \cite{levy1940certains,feller1950introduction}, 
\begin{equation*}
    \mathbb{P}[t_i(t)\leq \tau|x(0)=0]=\frac{2}{\pi} \arcsin\left(\sqrt{\frac{\tau}{t}} \right),
\end{equation*}
for $i\in\{m,r,\ell\}$. Then the probability densities of these quantities is given by,
\begin{equation}\label{eq:arcsine_law}
    \mathcal{P}(t_i,t|0)=\frac{1}{\pi \sqrt{t_i(t-t_i)}},
\end{equation}
which displays the counterintuitive property that the density diverges at $t_i=0$ and $t_i=t$, which means the value of $t_i$ is more likely to be either extreme, with the mean, $t/2$, being the minimum. 

The first Arcsine law in particular has gained a lot of interest due to its prominent role in the field of extreme value statistics \cite{gumbel1958statistics,majumdar2020extreme}, where one is often interested in the maximum position, $M(t)$, as well as the time of this event, $t_m(t)$. In this case the joint density is sought, which in the BM case is given by \cite{schehr2010extreme},
\begin{equation}\label{eq:free_max_double}
    \mathcal{P}(M,t_m,t|0)=\frac{M}{2 \pi D t_m^{3/2}\sqrt{t-t_m}}e^{\frac{-M^2}{4 D t_m}},
\end{equation}
where $D$ is the diffusion constant. Marginalization over $M$ of equation (\ref{eq:free_max_double}) recovers equation (\ref{eq:arcsine_law}) whereas over $t_m$, the following half-Gaussian distribution is found \cite{majumdar2020extreme}
\begin{equation}\label{eq:brownian_max}
    \mathcal{P}(M,t|0)=\frac{e^{-\frac{M^2}{4D t}}}{\sqrt{\pi D t}}.
\end{equation}

In recent literature there has been a large effort to extend these EVS and Arcsine laws to when the underlying stochastic process is not the simple BM. EVS studies include, various generalizations of BM \cite{schehr2010extreme,randon2007distribution,majumdar2008time,schehr2008exact,mori2021distribution,mori2022time,singh2021extremal,singh2022extreme,fyodorov2016moments,delorme2016extreme,delorme2016perturbative,sadhu2018generalized}, as well as other stochastic processes such as Bessel \cite{schehr2010extreme}, L\'{e}vy flights \cite{andersen1953fluctuations,majumdar2010universal}, random acceleration \cite{majumdar2010time,burkhardt2008first} and run-and-tumble \cite{singh2019generalised,mori2020universal,mori2020dimensional,singh2022extremal}. More recently, the time between the maximum and minimum of a stochastic process has been studied as well \cite{mori2019time,mori2020distribution}. In addition, the other two Arcsine laws have been studied, together or separately, in numerous contexts, such as constrained and resetting BM \cite{majumdar2008time,den2019properties}, BM in random mediums \cite{majumdar2002local}, bounded regions \cite {grebenkov2007residence, berezhkovskii1998residence,comtet2003brownian}, and with spatial and temporal heterogeneity \cite{singh2022extreme,bressloff2017residence}, as well as other stochastic processes such as continuous time random walks \cite{bel2005occupation}, random acceleration \cite{boutcheng2016occupation}, run-and-tumble \cite{singh2019generalised}, fractional BM \cite{sadhu2018generalized}, subdiffusion \cite{barkai2006residence,carmi2010distributions,korabel2011boundary} and random systems with quenched disorder \cite{burov2007occupation}. Despite this vast literature no such study on EVS and Arcsine laws of Brownian motion in the presence of permeable barriers has appeared. Here we provide the first such study.

The paper is structured as follows. In section \ref{sec:permeable} we introduce the key equations in modelling diffusion in the presence of permeable barriers. In section \ref{sec:max_max_time} we derive the joint density of $M(t)$ and $t_m(t)$ and study the marginal densities. Section \ref{sec:residence} and \ref{sec:last_passage} are devoted to the density of $t_r(t)$ and $t_\ell(t)$, respectively, while we summarize our work in section \ref{sec:conclusion}.

\section{Diffusion through Permeable Barriers}\label{sec:permeable}

We consider a Brownian particle, $x(t)$, undergoing diffusion in one dimension
in the presence of a permeable barrier at the origin. Classically, this has been
modelled by the diffusion equation (DE) along with the following permeable barrier
condition (PBC) \cite{tanner1978transient,powles1992exact}, 
\begin{equation}\label{eq:permeable_bc}
    J(0^\pm,t)=\kappa[P(0^-,t)-P(0^+,t)],
\end{equation}
where $P(x,t)$ represents the probability density of the position of the Brownian
particle at time $t$, with $J(x,t)=-D \partial_x P(x,t)$ being the probability current and the parameter $\kappa$ representing the permeability of the barrier, respectively, with $\kappa \to 0$ representing an impenetrable barrier (reflecting boundary) and no barrier when $\kappa\to \infty$. The notation $0^\pm$ indicates the right or left-hand side of the permeable barrier, respectively. 

As the DE with condition (\ref{eq:permeable_bc}) does not lend itself to study quantities other than $P(x,t)$ the present authors have introduced a new fundamental equation \cite{kay2022diffusion} with which one reformulates the problem in terms of a modified DE that accounts for the PBC in an inhomogeneous term (taken here at the origin for simplicity),
\begin{equation}\label{eq:diffusion_permeable}
    \frac{\partial P(x,t)}{\partial t}=D \frac{\partial^2 P(x,t)}{\partial x^2} + \frac{D}{\kappa}\delta'(x)J(0,t),
\end{equation}
where $\delta'(x)$ represents the derivative of the Dirac-$\delta$ function. The
convenience of equation (\ref{eq:diffusion_permeable}) is that for any initial condition localized at $x_0$, $P(x,0)=\delta(x-x_0)$ can be solved exactly in the Laplace domain as \cite{kay2022diffusion}
\begin{equation}\label{eq:diffusion_permeable_sol}
    \widetilde{P}(x,\epsilon|x_0)=\widetilde{G}_0(x,\epsilon|x_0)
    -\partial_{x_0}\widetilde{G}_0(x,\epsilon|0)\frac{\widetilde{J}_0(0,\epsilon|x_0)}{\frac{\kappa}{D}+ \partial_{x_0}\widetilde{J}_0(0,\epsilon|0)},
\end{equation}
in terms of the Green's function of Brownian motion in the absence of a permeable barrier, $G_0(x,t|x_0)=\exp\left\{-(x-x_0)^2/4Dt \right\}/\sqrt{4 \pi Dt}$. Generalizations to the case when an external potential is present are also possible with $G_0(x,t|x_0)$ becoming the Green's function of the associated Smoluchowski equation. In equation (\ref{eq:diffusion_permeable_sol}) we have used the Laplace variable $\epsilon$ to indicate that for an arbitrary time-dependent function, $f(t)$, has its Laplace transformed expression given by $\widetilde{f}(\epsilon)=\int_0^\infty e^{-\epsilon t}f(t)dt$. The notation $P(x,t|x_0)$ indicates the localized initial condition at $x_0$ and $J_0(x,t|x_0)=-D\partial_x G_0(x,t|x_0)$ is defined as the free probability current where $\partial_{x_0}\widetilde{G}_0(x,\epsilon|0)$ implies $\frac{\partial}{\partial x_0}\widetilde{G}_0(x,\epsilon|x_0)|_{x_0=0}$. 

Further convenience in employing the formalism associated with equation (\ref{eq:diffusion_permeable}) is due to the ability to write an equivalent Fokker-Planck (FP) representation, namely the homogeneous (It\^{o}) FP equation \cite{kay2022diffusion}, $\partial_t P(x,t)=L_x P(x,t)$ where $L_x=-\partial_x A(x) +\partial_x^2 B(x)$ is the Fokker-Planck operator, with the spatially dependent drift and diffusion coefficients given by $A(x)=-(D^2/\kappa)\delta'(x)$ and $B(x)=D-(D^2/\kappa)\delta(x)$. As will become apparent later, we are mainly interested in the backward FP (Kolmogorov) equation in terms of the initial variables, $t_0$ and $x_0$, $-\partial_{t_0} P(x_0,t_0)=L_{x_0}^\dag P(x_0,t_0)$, where $L^\dag_x$ is the formal adjoint of the Fokker-Planck operator i.e. $L^\dag_x=A(x)\partial_x+B(x) \partial_x^2 $. For $P(x_0,t_0=0)=\delta(x-x_0)$ and exploiting the time homogeneity of the process, we have \cite{risken1996fokker}
\begin{equation}\label{eq:backward_fp_equation}
    \frac{\partial}{\partial t}P(x,t|x_0)=A(x_0)\frac{\partial}{\partial x_0}P(x,t|x_0)+B(x_0)\frac{\partial^2}{\partial x_0^2} P(x,t|x_0).
\end{equation}
As was shown in Ref.
\cite{kay2022diffusion}, $L$ is in fact self-adjoint, such that the backward FP
equation, is given by 
\begin{equation}\label{eq:backwards_fp_permeable}
    \frac{\partial P(x,t|x_0)}{\partial t}=D \frac{\partial^2 P(x,t|x_0)}{\partial x_0^2}-\frac{D^2}{\kappa} \delta'(x_0)\partial_{x_0}P(x,t|0),
\end{equation}
which implies that we can solve equation (\ref{eq:backwards_fp_permeable}) through the same procedure for which we obtain equation (\ref{eq:diffusion_permeable_sol}).

\section{Extreme value $M$ and Time to Reach Maximum $t_m$}\label{sec:max_max_time}

\subsection{Joint Probability Density $\mathcal{P}(M,t_m,t|0^\pm)$}

We study the time-dependent joint distribution, $\mathcal{P}(M,t_m,t|x_0)$, of the maximum position $M=x(t_m)$ and the time $t_m$ at which this occurs for a Brownian particle in the presence of a permeable barrier at the origin. Since we consider the particle starting from $x_0=0$ the presence of the permeable barrier leads to two different joint densities $\mathcal{P}(M,t_m,t|0^+)$ and $\mathcal{P}(M,t_m,t|0^-)$. 

To find these joint densities we proceed by using a path decomposition approach \cite{majumdar2008time,majumdar2010time,singh2022extreme,mori2022time}, where we exploit the Markovian nature of the process to split the trajectories of the Brownian particle into two parts. The first part is $\{x(\tau): \tau \in [0,t_m]\}$ and the second part is $\{x(\tau): \tau \in [t_m,t] \}$, see figure \ref{fig:trajectory}. Clearly the first part of the trajectory can be expressed as the probability of reaching $M$ for the first time at $t_m$, which is the first-passage time distribution $\mathcal{F}(M,t_m|0^\pm)$. The second part is the probability that the particle starting at $M$ does not reach this position again in the remaining time, this is the survival probability $S(M,t-t_m|M)$. As $S(M,t-t_m|M)=0$, we remedy this by using the quantity $S(M+\varepsilon,t-t_m|M)$ and taking $\varepsilon\to 0^+$ \cite{majumdar2008time,singh2022extreme}. 

We now proceed to find the two quantities, $\mathcal{F}(x,t|x_0)$ and $S(x,t|x_0)$. This calculation was performed in detail in Ref. \cite{kay2022diffusion}, where using equation (\ref{eq:backwards_fp_permeable}) a backward equation was found for $S(x,t|x_0)$, which was used to find $\mathcal{F}(x,t|x_0)$ in the Laplace domain through $\widetilde{\mathcal{F}}(x,\epsilon|x_0)=1-\epsilon \widetilde{S}(x,\epsilon|x_0)$:
\begin{equation}\label{eq:fpt}
    \fl \widetilde{\mathcal{F}}(x,\epsilon|x_0)=\left\{
    \begin{array}{ll}
        \frac{2\kappa e^{-|x-x_0|\sqrt{\frac{\epsilon}{D}}}}{\sqrt{D\epsilon}\left[1+e^{-2|x|\sqrt{\frac{\epsilon}{D}}}\right]+2\kappa}, \ x_0<0<x \ \ \mathrm{or} \ \ x<0<x_0,\\
        \frac{\left[\sqrt{D \epsilon}+2\kappa\right] e^{-|x-x_0|\sqrt{\frac{\epsilon}{D}}}+\sqrt{D \epsilon}e^{-|x+x_0|\sqrt{\frac{\epsilon}{D}}}}{\sqrt{D\epsilon}\left[1+e^{-2|x|\sqrt{\frac{\epsilon}{D}}}\right]+2\kappa}, \ 0<x_0<x \ \ \mathrm{or} \ \ x<x_0<0.
    \end{array}
    \right .
\end{equation}
We now proceed to calculate the joint distributions for $x_0=0^\pm$.

\subsubsection*{Case $x_0=0^+$.}
\hfill \break
\hfill \break
The joint distribution is given by 
\begin{equation}\label{eq:joint_plus}
    \mathcal{P}(M,t_m,t|0^+)= \lim_{\varepsilon\to 0^+} \frac{\mathcal{F}(M,t_m|0^+)S(M+\varepsilon,t-t_m|M)}{N(\varepsilon)},
\end{equation}
where $N(\varepsilon)$ ensures the quantity is normalized, alternatively after Laplace transforming with respect to $t_m$ and $t$ ($t_m\to p$ and $t\to \epsilon$) we have, 
\begin{eqnarray}\label{eq:joint_plus_laplace}
    &\widetilde{Q}_m(M,p,\epsilon|0^+)=\lim_{\varepsilon\to 0^+} \frac{1}{N(\varepsilon)} \int_{0}^{\infty} d t_m e^{-p t_m} \mathcal{F}(M,t_m|0^+) \\ \nonumber
    &\times \int_{0}^{\infty} d t e^{-\epsilon t} S(M+\varepsilon,t-t_m|M) 
    = \lim_{\varepsilon\to 0^+} \frac{\widetilde{\mathcal{F}}(M,p+\epsilon|0^+)\widetilde{S}(M+\varepsilon,\epsilon|M)}{N(\varepsilon)},
\end{eqnarray}
where we have used the notation $Q_m(M,p,t|0^\pm)=\int_0^\infty dt_m e^{-pt_m} \mathcal{P}(M,t_m,t|0^\pm)$. 

Expanding $\widetilde{S}(M+\varepsilon,\epsilon|M)$ to first order in $\varepsilon$, we have $\widetilde{S}(M+\varepsilon,\epsilon|M)\simeq -\varepsilon\lim_{x\to M}\partial_{x}\widetilde{\mathcal{F}}(x,\epsilon|M)/\epsilon$ and requiring $\mathcal{P}(M,t_m,t|0^\pm)$ to be normalized, namely, $\int_0^\infty \widetilde{Q}_m(M,0,\epsilon|0^\pm)dM=\epsilon^{-1}$, we find $N(\varepsilon)=\varepsilon$, and then obtain
\begin{eqnarray}\label{eq:max_plus_double_laplace}
    \fl \widetilde{Q}_m(M,p,\epsilon|0^+)\\ \nonumber
    \fl=\frac{2 e^{M \sqrt{\frac{p+\epsilon }{D}}} \left(e^{2 M \sqrt{\frac{\epsilon }{D}}} \left(\sqrt{D \epsilon }+2 \kappa \right)-\sqrt{D \epsilon }\right) \left(\sqrt{D (p+\epsilon )}+\kappa \right)}{\sqrt{D \epsilon } \left(e^{2 M \sqrt{\frac{\epsilon }{D}}} \left(\sqrt{D \epsilon }+2 \kappa \right)+\sqrt{D \epsilon }\right) \left(e^{2 M \sqrt{\frac{p+\epsilon }{D}}} \left(\sqrt{D (p+\epsilon )}+2 \kappa \right)+\sqrt{D (p+\epsilon )}\right)}.
\end{eqnarray}
One can see from equation (\ref{eq:max_plus_double_laplace}) that in the no barrier limit, $\kappa\to\infty$, equation (\ref{eq:max_plus_double_laplace}) correctly reduces to $\widetilde{Q}_m(M,p,\epsilon|0)=e^{-M \sqrt{(p+\epsilon)/D}}/\sqrt{D \epsilon }$, giving equation (\ref{eq:free_max_double}) after the double Laplace inversion.

Similarly, in the limit of the barrier becoming impermeable, $\kappa\to 0$, we find, 
\begin{equation}\label{eq:reflect_max_laplace}
    \widetilde{Q}_m(M,p,\epsilon|0^+)=\frac{\tanh \left(M \sqrt{\frac{\epsilon} {D}}\right) \mathrm{sech}\left(M \sqrt{\frac{p+\epsilon}{D }}\right)}{\sqrt{D \epsilon}}
\end{equation}
after using the following inverse Laplace transform relations: $\mathcal{L}^{-1}_{\epsilon\to t}\left\{ \epsilon^{-1/2} \tanh(a\sqrt{\epsilon}) \right\}= \vartheta_4(0,e^{-a^2/t})/\sqrt{\pi t}$ and $\mathcal{L}^{-1}_{\epsilon\to t}\left\{\mathrm{sech}(a\sqrt{\epsilon}) \right\}=a\vartheta_1'(0,e^{-a^2/t})/\sqrt{4 \pi t^3}$ \cite{roberts1966table}, equation (\ref{eq:reflect_max_laplace}) becomes,
\begin{equation}\label{eq:reflect_joint_max}
    \mathcal{P}(M,t_m,t|0^+)=\frac{M \vartheta_1'\left(0,e^{-\frac{M^2}{D t_m}}\right)\vartheta_4\left(0,e^{-\frac{M^2}{D (t-t_m)}}\right)}{2 \pi D t_m^{3/2}\sqrt{t-t_m}},
\end{equation}
where $\vartheta_1(z,q)=\sum_{n=-\infty}^{\infty}(-1)^{n-1/2}q^{(n+\frac{1}{2})^2}e^{(2n+1)iz}$ and $\vartheta_4(z,q)=\sum_{n=-\infty}^{\infty}(-1)^n q^{n^2}e^{2ni z} $ are the Jacobi Theta functions and $\vartheta_n'(z,q)$ represents the derivative with respect to $z$ \cite{abramowitz1988handbook}. 

Interestingly, if we take the small $t$ limit of $\mathcal{P}(M,t_m,t|0^+)$, which corresponds to $\epsilon,p \to \infty$ in the Laplace domain for equation (\ref{eq:max_plus_double_laplace}), we obtain to leading order equation (\ref{eq:reflect_max_laplace}). This implies that for $t\to 0$ the permeable barrier acts as if it is fully reflecting. This can be understood by considering for very small times the particle does not have enough time to interact with the barrier and pass through, meaning that essentially no trajectories reach the other side of the barrier.

Although the double inverse Laplace transform of equation (\ref{eq:max_plus_double_laplace}) for arbitrary $\kappa$ looks highly non-trivial, significant ground can be made (see \ref{sec:joint_max_inverse_laplace}). We find $\mathcal{P}(M,t_m,t|0^+)$ in terms of two different scaling functions such that,
\begin{equation}\label{eq:joint_max}
    \mathcal{P}(M,t_m,t|0^+)=\frac{\kappa^3}{D^2}\mathdutchcal{G}^+\left(\frac{\kappa}{D}M,\frac{\kappa^2}{D} t_m\right) \mathdutchcal{H}\left(\frac{\kappa}{D}M,\frac{\kappa^2}{D} (t-t_m)\right),
\end{equation}
where,
\begin{equation}\label{eq:g_plus}
    \mathdutchcal{G}^+(y,\tau_1)=\frac{1}{\pi} \int_0^\infty e^{-\tau_1 z} \sin(y \sqrt{z}) h(y,z) dz,
\end{equation}
and 
\begin{equation}\label{eq:max_H}
    \mathdutchcal{H}(y,\tau_2)=\frac{4}{\pi} \int_0^\infty \frac{e^{-\tau_2 z}}{\sqrt{z}}h(y,z) dz,
\end{equation}
with 
\begin{equation}\label{eq:small_h}
    h(y,z)=\left[2+z+z\cos(2y\sqrt{z})+2\sqrt{z}\sin(2y\sqrt{z})\right]^{-1}.
\end{equation}
From equation (\ref{eq:joint_max}) one can see we no longer have the transformation symmetry of $t_m\to t-t_m$, meaning the symmetry about $t/2$ that one observes in the barrier free case, equation (\ref{eq:free_max_double}), is broken.

\subsubsection*{Case $x_0=0^-$.}
\hfill \break
\hfill \break
For $x_0=0^-$ we have the added complexity due to the chance that the particle will not cross the barrier throughout the whole time period, leading to the maximum occurring at the initial position $x_0$. The probability of this occurring is given by (from equation (\ref{eq:fpt}))
\begin{equation}\label{eq:prob_no_cross}
    S(0^\pm,t|0^\mp)=e^{\frac{\kappa^2t}{D}} \mathrm{erfc}\left(\kappa\sqrt{\frac{t}{D}}\right),
\end{equation}
where $\mathrm{erfc}(z)=1-\mathrm{erf}(z)$ where $\mathrm{erf}(z)=(2/\sqrt{\pi})\int_0^z du e^{-u^2}$. This means that the maximum position is the origin and it occurs at time $t_m=0$. Then from equation (\ref{eq:joint_plus}) we have,
\begin{equation}\label{eq:joint_minus}
    \fl \mathcal{P}(M,t_m,t|0^-)= \lim_{\varepsilon\to 0^+} \frac{1}{N(\varepsilon)} \big[\mathcal{F}(M,t_m|0^-)S(M+\varepsilon,t-t_m|M)\\
    +S(0^+,t|0^-)\delta(M)\delta(t_m)\big].
\end{equation}
Similar to equation (\ref{eq:joint_plus_laplace}) we perform a double Laplace transform and obtain
\begin{equation}\label{eq:joint_minus_laplace}
    \fl \widetilde{Q}_m(M,p,\epsilon|0^-)= \lim_{\varepsilon\to 0^+} \frac{1}{N(\varepsilon)}\big[\widetilde{\mathcal{F}}(M,p+\epsilon|0^-)\widetilde{S}(M+\varepsilon,\epsilon|M) +\widetilde{S}(0^+,\epsilon|0^-)\delta(M)\big].
\end{equation}
Expanding to first order in $\varepsilon$ and ensuring normalization we once again find $N(\varepsilon)=\varepsilon$, leading to
\begin{equation}\label{eq:max_minus_double_laplace}
    \widetilde{Q}_m(M,p,\epsilon|0^-)=\frac{\kappa}{\sqrt{D(p+\epsilon)}+\kappa}\widetilde{Q}_m(M,p,\epsilon|0^+) +\frac{D\delta(M)}{\kappa  \sqrt{D \epsilon }+D \epsilon }.
\end{equation}
As in equation (\ref{eq:max_plus_double_laplace}) one can see in the limit $\kappa\to \infty$ equation (\ref{eq:max_minus_double_laplace}) reduces to the barrier free case, equation (\ref{eq:free_max_double}). In the fully reflecting limit $\kappa \to 0$ one recovers $\mathcal{P}(M,t_m,t|0^-)=\delta(M)\delta(t_m)$, since no particles pass through, meaning the maximum position will be at the origin being reached instantly. As in the $x_0=0^+$ case, we see from equation (\ref{eq:max_minus_double_laplace}) that in short time limit one recovers the fully reflecting limit up to leading order in $t_m$. However, using $\widetilde{Q}_m(M,p,\epsilon|0^+)$ for $p,\epsilon \to \infty$, equation (\ref{eq:reflect_max_laplace}), we find a  better approximation
\begin{equation}\label{eq:joint_min_reflect_laplace}
    \fl \widetilde{Q}_m(M,p,\epsilon|0^-)\simeq \frac{\kappa\tanh \left(M \sqrt{\frac{\epsilon} {D}}\right) \mathrm{sech}\left(M \sqrt{\frac{p+\epsilon}{D }}\right)}{D\sqrt{(p+\epsilon)\epsilon}}+\left(\frac{1}{\epsilon}-\frac{\kappa}{\sqrt{D \epsilon^3}}  \right)\delta(M),
\end{equation}
which after using $\mathcal{L}^{-1}_{\epsilon\to t}\left\{ \epsilon^{-1/2} \mathrm{sech}(a\sqrt{\epsilon}) \right\}=\theta(e^{-a^2/t})/\sqrt{\pi t}$ \cite{roberts1966table}, equation (\ref{eq:joint_min_reflect_laplace}) becomes
\begin{equation}\label{eq:joint_max_minus_approx}
    \fl \mathcal{P}(M,t_m,t|0^-)\simeq \frac{\kappa \theta \left(e^{-\frac{M^2}{D t_m}}\right) \vartheta_4\left(0,e^{-\frac{M^2}{D (t-t_m)}}\right)}{\pi D \sqrt{t_m(t-t_m)}}+\left(1-\frac{2\kappa\sqrt{t}}{\sqrt{\pi D}}\right)\delta(M)\delta(t_m),
\end{equation}
where $\theta(q)=\sum_{n=0}^{\infty}(-1)^nq^{(n+\frac{1}{2})^2}$.

Now we perform the double inverse Laplace transform of Eq. (\ref{eq:max_minus_double_laplace}) to obtain (see \ref{sec:joint_max_inverse_laplace})
\begin{eqnarray}\label{eq:joint_min}
    \mathcal{P}(M,t_m,t|0^-)&=\frac{\kappa^3}{D^2}\mathdutchcal{G}^-\left(\frac{\kappa}{D}M,\frac{\kappa^2}{D} t_m\right) \mathdutchcal{H}\left(\frac{\kappa}{D}M,\frac{\kappa^2}{D} (t-t_m)\right)\\\nonumber
    &+\frac{\kappa^3}{D^2}\mathdutchcal{I}\left(\frac{\kappa}{D}M,\frac{\kappa^2}{D} t_m,\frac{\kappa^2}{D} (t-t_m)\right)
\end{eqnarray}
where $\mathdutchcal{I}(y,\tau_1,\tau_2)=e^{\tau_2}\mathrm{erfc}(\sqrt{\tau_2})\delta(y)\delta(\tau_1)$, $\mathdutchcal{H}(y,\tau_2)$ is defined in equation (\ref{eq:max_H}) and 
\begin{equation}\label{eq:g_minus}
    \mathdutchcal{G}^-(y,\tau_1)= \int_0^\infty \frac{e^{-\tau_1 z}}{\pi} \left(\sin(y \sqrt{z})+\sqrt{z}\cos(y \sqrt{z})\right) h(y,z) dz,
\end{equation}
where $h(y,z)$ is defined in equation (\ref{eq:small_h}). Again we see the $t_m\to t-t_m$ symmetry breaking as in the $x_0=0^+$ case. 

\subsection{Marginal Density $\mathcal{P}(M,t|0^\pm)$}

To find the marginal density $\mathcal{P}(M,t|0^\pm)$ one integrates over $t_m$, i.e. $p\to0$ in equations (\ref{eq:max_plus_double_laplace}) and (\ref{eq:max_minus_double_laplace}) and after taking the double Laplace inverse (see \ref{sec:max_inverse_laplace}) we get
\begin{equation}\label{eq:max_marginal}
    \mathcal{P}(M,t|0^\pm)=\frac{\kappa}{D}\mathcal{I}^\pm\left(\frac{\kappa}{D}M,\frac{\kappa^2}{D}t\right)
\end{equation}
where 
\begin{equation}\label{eq:h_plus}
    \mathcal{I}^+(y,\tau)=\frac{1}{\pi}\int_0^\infty \frac{e^{-\tau z}}{\sqrt{z}}j^+(y,z)h^2(y,z)dz,
\end{equation}
and 
\begin{equation}\label{eq:h_minus}
    \mathcal{I}^-(y,\tau)=\frac{1}{\pi}\int_0^\infty \frac{e^{-\tau z}}{\sqrt{z}}j^-(y,z)h^2(y,z)dz + e^\tau\mathrm{erfc}(\sqrt{\tau})\delta(y),
\end{equation}
with $j^\pm(y,z)$ defined in equations (\ref{eq:j_plus}) and (\ref{eq:j_minus}) and $h(y,z)$ defined in equation (\ref{eq:small_h}). We have verified that in the barrier free case ($\kappa \to \infty$) one recovers equation (\ref{eq:brownian_max}).

From equations (\ref{eq:max_marginal}), (\ref{eq:h_plus}) and (\ref{eq:h_minus}), we find that for $t>>D/\kappa^2$ with $M\sim D/\kappa$, the integrands in (\ref{eq:h_plus}) and (\ref{eq:h_minus}) are dominated by small $z$, so we expand $j^\pm(y,z)h^2(y,z)$ to first order in $z$ and then compute the integral to obtain,
\begin{equation}\label{eq:max_long_time_plus}
    \mathcal{P}(M,t|0^+)\simeq \frac{4 \kappa D (\kappa t-2M)-\kappa^2 M^2-2D^2}{4 \sqrt{\pi}\kappa^2 (Dt)^{3/2}}
\end{equation}
and 
\begin{equation}\label{eq:max_long_time_minus}
    \mathcal{P}(M,t|0^-)\simeq \frac{4 \kappa D (\kappa t-\frac{5}{2}M)-\kappa^2 M^2-6D^2}{4 \sqrt{\pi}\kappa^2 (Dt)^{3/2}}+\frac{\delta(M)}{\kappa\sqrt{\pi D t}}.
\end{equation}
Equations (\ref{eq:max_long_time_plus}) and (\ref{eq:max_long_time_minus}) should be compared with the barrier free case, $\mathcal{P}(M,t|0)\simeq (Dt-M^2/4)/\sqrt{\pi D^3 t^3}$.

\begin{figure}[htp]
    \centering
    \subfloat{\includegraphics[width=0.5\textwidth]{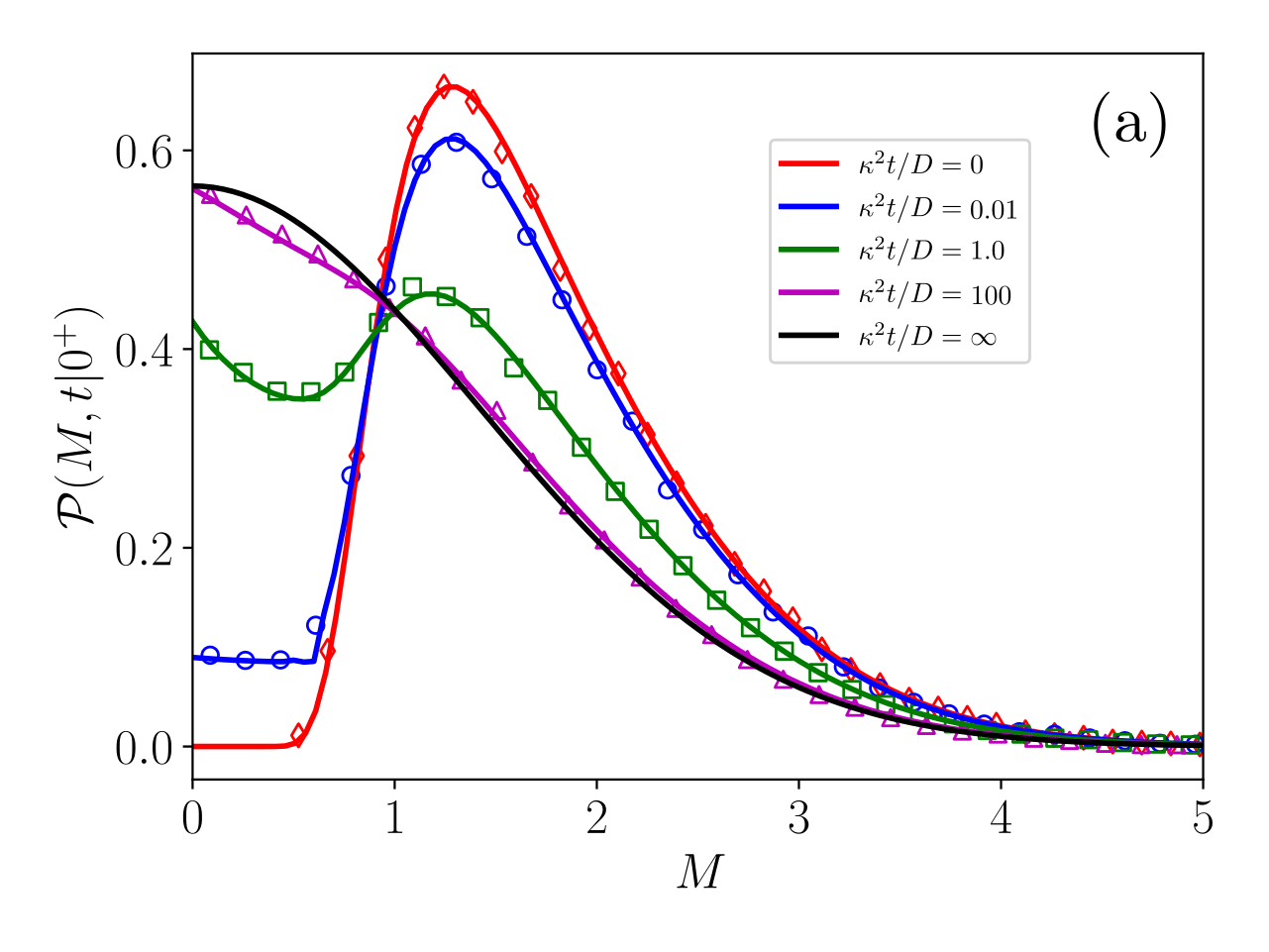}}
    \subfloat{\includegraphics[width=0.5\textwidth]{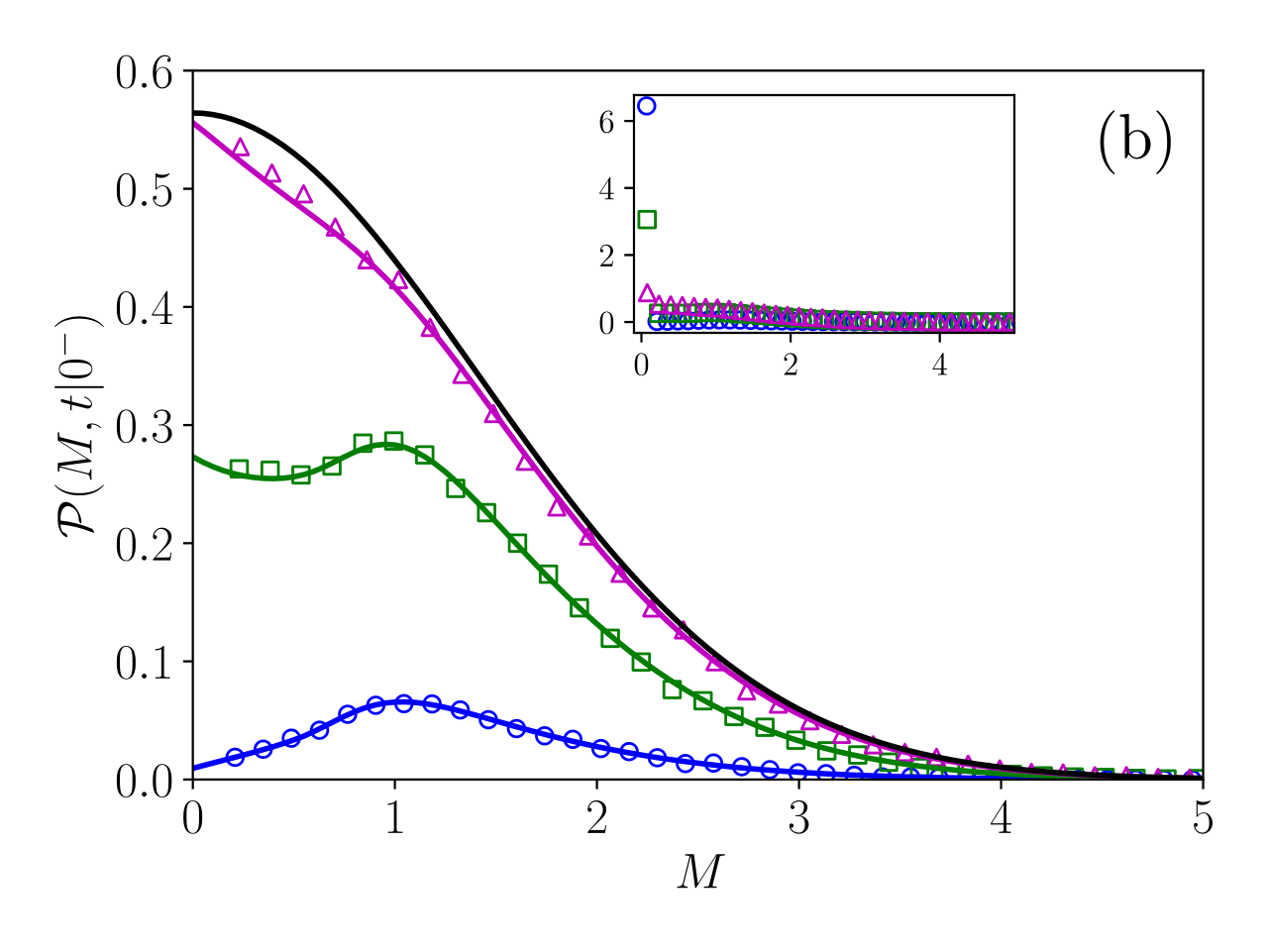}}
    \caption{Maximum position density, $\mathcal{P}(M,t|0^\pm)$, where a permeable barrier has been placed at the origin. Two distinct initial positions are considered, (a) $x_0=0^+$ and (b) $x_0=0^-$, for different values of the dimensionless parameter $\kappa^2 t/D$, all other quantities are in arbitrary units. Markers represent stochastic simulations. The inset in figure (b) shows the presence of the Dirac-$\delta$ from simulations (the analytic Dirac-$\delta$ is omitted from the plot).}
    \label{fig:max}
\end{figure}

We plot equation (\ref{eq:max_marginal}) in figure \ref{fig:max} for the two initial conditions, $x_0=0^\pm$ whilst varying the dimensionless parameter, $\kappa^2 t/D$, and compare to stochastic simulations. One can see for certain parameter values, namely for small permeabilities, the distributions become non-monotonic, and we have a bi-modal shape. In the $x_0=0^+$ case this feature can be explained by considering for small $\kappa$, the particle is unlikely to pass through the barrier and thus being more likely to move away from the barrier leading to a maximum occurring for $M>0$. Whereas the peak at the origin is caused by the particle managing to pass through the barrier but never returning. The presence of a local minimum near $M=0$ is thus caused by the less likely scenario in which the particle stays near the barrier, constantly interacting with it, but without venturing too far from the origin. The case $x_0=0^-$ can be explained similarly except there is a non-zero probability that the particle never crosses the barrier leading to the global maximum always occurring at the origin (the Dirac-$\delta$ function).


\subsection{Marginal Density $\mathcal{P}(t_m,t|0^\pm)$}

Here we consider the marginal density $\mathcal{P}(t_m,t|0^\pm)=\int_0^\infty \mathcal{P}(M,t_m,t|0^\pm)dM$, where from equations (\ref{eq:joint_max}) and (\ref{eq:joint_min}), we obtain the scaling relation,
\begin{equation}\label{eq:max_time_density}
    \mathcal{P}(t_m,t|0^\pm)=\frac{\kappa^2}{D}\mathdutchcal{F}^\pm_m\left(\frac{\kappa^2}{D}t_m,\frac{\kappa^2}{D}(t-t_m)\right),
\end{equation}
where 
\begin{equation}\label{eq:max_time_scaling}
    \mathdutchcal{F}^\pm_m(\tau_1,\tau_2)=\int_0^\infty \mathdutchcal{G}^\pm(y,\tau_1)\mathdutchcal{H}(y,\tau_2)dy.
\end{equation}

The integral in equation (\ref{eq:max_time_scaling}) is hard to compute, but we can study the asymptotic forms of $\mathdutchcal{F}^\pm_m(\tau_1,\tau_2)$. Firstly, we study the short time asymptotics, $t<<D/\kappa^2$, which can be approximated by the marginal over $M$ of equations (\ref{eq:reflect_joint_max}) and (\ref{eq:joint_max_minus_approx}). By using the series form of the Jacobi theta functions and integrating over we have
\begin{eqnarray}
    \mathdutchcal{F}_m^+(\tau_1,\tau_2)&\simeq \sum_{n=-\infty}^\infty \sum_{m=-\infty}^\infty \frac{(-1)^{m+n} (2 n+1) \sqrt{\tau_2} }{\pi  \sqrt{\tau_1} \left(4 m^2 \tau_1+(2 n+1)^2 \tau_2\right)} \\ \nonumber
    &= \sum _{n=-\infty }^{\infty } \frac{(-1)^n \mathrm{cosech} \left((n+\frac{1}{2})\pi \sqrt{\frac{\tau_2}{\tau_1}}\right)}{2 \tau_1},
\end{eqnarray}
where we used $\sum_{m=-\infty}^{\infty} (-1)^m (m^2+z)^{-1}=\pi \mathrm{cosech}(\pi \sqrt{z})z^{-1/2}$. And,
\begin{equation}
    \mathdutchcal{F}_m^-(\tau_1,\tau_2)\simeq  \sum_{n=-\infty}^\infty \sum_{m=-\infty}^\infty (-1)^{m+n}(4 \pi  m^2 \tau_1+(2 n+1)^2\pi \tau_2)^{-1/2}.
\end{equation}

Now we study the long time asymptotics of $\mathcal{P}(t_m,t|0^\pm)$, which corresponds to $t>>D/\kappa^2$. We find it more convenient to do this in the Laplace domain, where we use equations (\ref{eq:max_plus_double_laplace}) and (\ref{eq:max_minus_double_laplace}). There are two regimes which give the full picture of the asymptotics for $t\to \infty$: keeping $t_m$ finite in the first regime and keeping $t-t_m$ finite in the second. This corresponds to $\epsilon\to 0$ with $p >> \epsilon$ and $p \sim \epsilon$ in the Laplace domain, respectively. Expanding around $\epsilon,p\to0$ to leading order in equations (\ref{eq:max_plus_double_laplace}) and (\ref{eq:max_minus_double_laplace}) and integrating over $M$ one can see that we obtain, $Q_m(p,\epsilon|0^\pm)=1/\sqrt{\epsilon(p+\epsilon)}$, which after double inverse Laplace transforming recovers the Arcsine distribution, (\ref{eq:arcsine_law}). For the case $p>>\epsilon$, let us expand $\widetilde{Q}_m(p,\epsilon|0^+)$ around $\epsilon\to 0$ keeping $p$ finite, then we find
\begin{equation}
    \widetilde{Q}_m(p,\epsilon|0^+)\simeq \frac{2 (\sqrt{D p}+\kappa)}{\sqrt{D \epsilon}}\int_0^\infty \frac{e^{-M \sqrt{\frac{p}{D}}}}{\sqrt{D p}\left(1+e^{-M \sqrt{\frac{p}{D}}}\right)+2 \kappa}dM,
\end{equation}
which gives after performing the integral and inverse Laplace transforming with respect to $\epsilon$,
\begin{equation}\label{eq:arctan_laplace}
    Q_m(p,t|0^+)\simeq \frac{2 \left(\sqrt{D p}+\kappa \right) \mathrm{arctan}\left(\frac{(D p)^{1/4}}{(\sqrt{D p}+2 \kappa )^{1/2}}\right)}{\sqrt{\pi t}(D p)^{1/4} \left[p   \left(\sqrt{D p}+2 \kappa \right)\right]^{1/2}}.
\end{equation}
Since equation (\ref{eq:arctan_laplace}) is very difficult invert, we can investigate the asymptotic dependence for $t_m\to 0$. This corresponds to $p\to \infty$, thus by expanding Eq. (\ref{eq:arctan_laplace}) to leading order for $p\to \infty$ and using $\mathrm{arctan}(1)=\pi/4$, we find $Q_m(p,t|0^+)\sim \frac{\sqrt{\pi} }{2 \sqrt{p t }}$, and performing the Laplace inversion with respect to $p$ we get,
\begin{equation}\label{eq:max_plus_time_small}
    \mathcal{P}(t_m,t|0^+)\sim \frac{1}{2 \sqrt{t_m t}} \ \mathrm{for} \ t_m\to 0 ,\ t\to \infty.
\end{equation}

For the case $x_0=0^-$, using equation (\ref{eq:max_minus_double_laplace}), we find,
\begin{equation}\label{eq:max_minus_time_small}
    \mathcal{P}(t_m,t|0^-)\sim \frac{\sqrt{\pi}\kappa}{2 \sqrt{D t}} + \frac{\sqrt{D}}{\kappa \sqrt{\pi t}} \delta(t_m) \ \mathrm{for} \ t_m\to 0 ,\ t\to \infty.
\end{equation}
It is clear that equations (\ref{eq:max_plus_time_small}) and (\ref{eq:max_minus_time_small}) deviate from the Arcsine law, $\sim 1/\pi(t_m t)^{-1/2}$, showing the permeable barrier has an influence for $t\to \infty$ when $t_m$ is finite. And for the $x_0=0^-$ case the singularity at $t_m\to0$ is provided by the Dirac-$\delta$ function.

\begin{figure}[htp]
    \centering
    \subfloat{\includegraphics[width=0.5\textwidth]{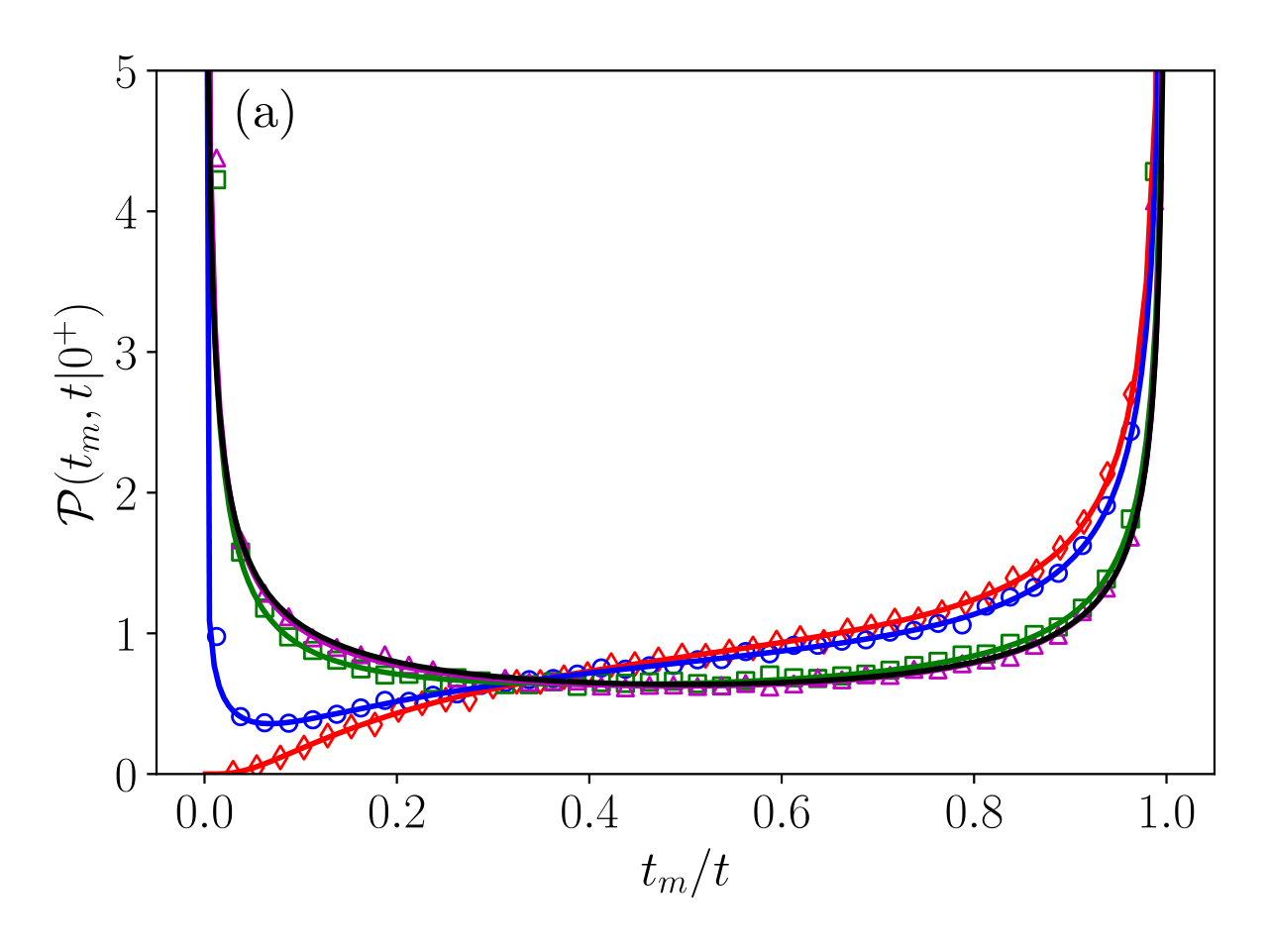}}
    \subfloat{\includegraphics[width=0.5\textwidth]{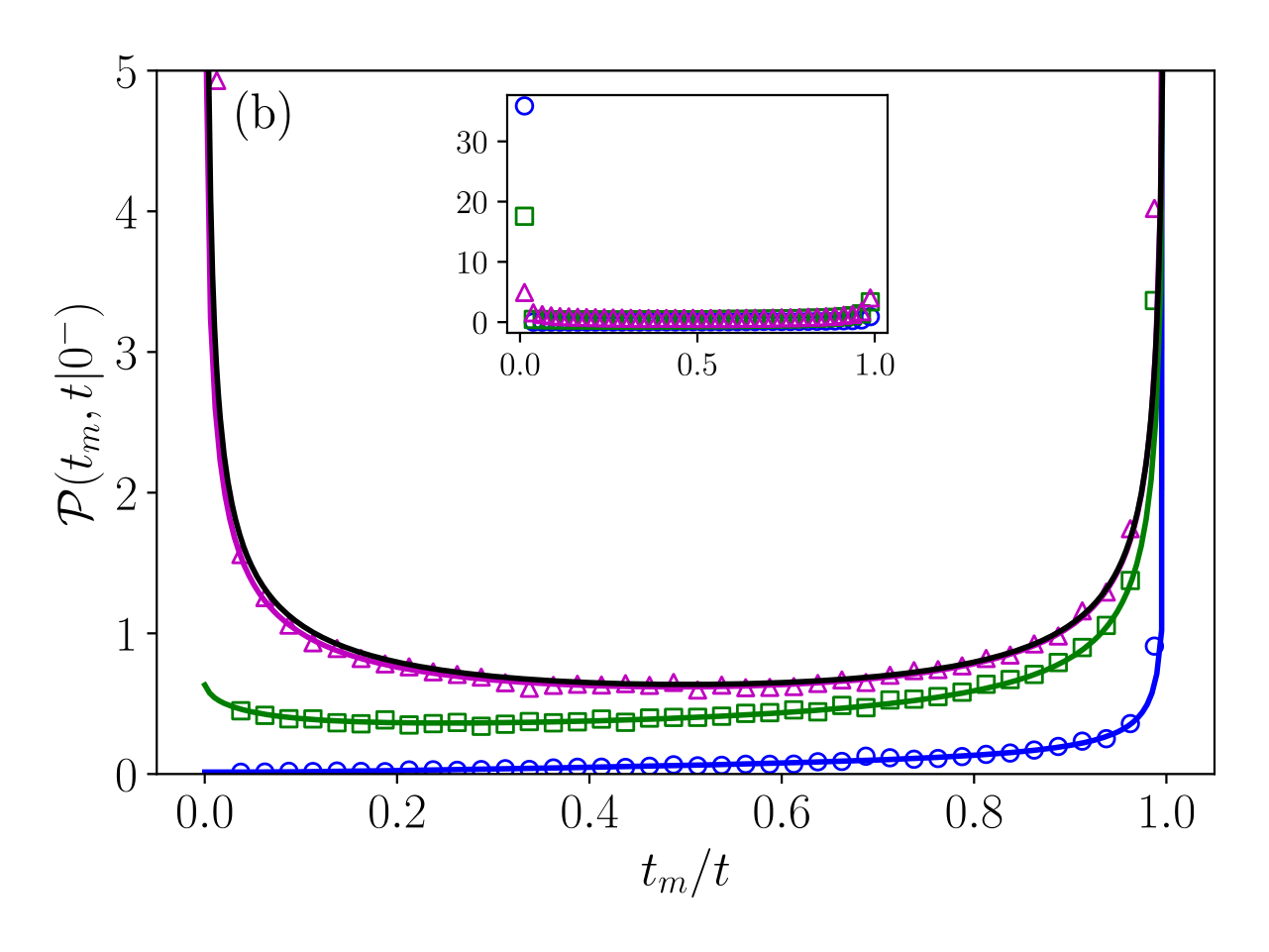}}
    \caption{Time of maximum position density, $\mathcal{P}(t_m,t|0^\pm)$, where a permeable barrier has been placed at the origin. Two distinct initial positions are considered, (a) $x_0=0^+$ and (b) $x_0=0^-$, for different values of the dimensionless parameter $\kappa^2 t/D$, all other quantities are in arbitrary units. Markers represent stochastic simulations. The inset in figure (b) shows the presence of the Dirac-$\delta$ from simulations (the analytic Dirac-$\delta$ is omitted from the plot). The parameter values for each curve are the same as in figure \ref{fig:max} (a).} 
    \label{fig:max_time}
\end{figure}

We plot equation (\ref{eq:max_time_density}) in figure \ref{fig:max_time} to show the excellent match to simulations. One can see that for the different initial conditions $x_0=0^\pm$, $\mathcal{P}(t_m,t|0^\pm)$ has two different shapes. When $x_0=0^+$ and for small permeabilities one can see the minimum of the distribution is just after $t_m=0$, which is due to the small likelihood of the particle reaching its maximum and then passing through the barrier and never returning. The peak at $t_m=0$ indicates the particle instantly crosses the barrier and never returns. For $x_0=0^-$ and for small permeabilities the distribution is strongly weighted by the Dirac-$\delta$ function indicating the particle never crosses the barrier. However, if the barrier is crossed, it is very unlikely for the particle to cross again meaning it is more likely to reach the maximum at $t_m=t$.

\section{Residence Time in Positive Half-Space $t_r$}\label{sec:residence}

Here we study the second Arcsine law, namely the residence time the particle, starting at the origin, spends in the region $x>0$ when a permeable barrier is placed at the origin. Once again, due to the presence of this permeable barrier, we have two distinct initial positions, $x_0=0^+$ and $x_0=0^-$. To proceed we utilize the fact that the residence time can be written as the functional, $t_r(t)=\int_0^t \Theta(x(t'))dt'$, where $\Theta(z)$ is the Heaviside step function. From Feynman-Kac theory \cite{kac1949distributions,majumdar2007brownian}, we know that $Q_r(p,t|x_0)=\int_0^\infty e^{-p t_r} \mathcal{P}(t_r,t|x_0)dt_r=\left\langle e^{-p t_r(t)}\Big| x_0=x(0)\right\rangle$ satisfies the following backward Feynman-Kac equation, 
\begin{equation}\label{eq:backward_fk}
    \fl \frac{\partial Q_r(p,t|x_0)}{\partial t}=A(x_0)\frac{\partial Q_r(p,t|x_0)}{\partial x_0}+B(x_0)\frac{\partial^2 Q_r(p,t|x_0)}{\partial x_0^2}
    -p \Theta(x_0)Q_r(p,t|x_0),
\end{equation}
where we use the backward Fokker-Planck operator in equation (\ref{eq:backward_fp_equation}). Using the self-adjoint nature of this backward Fokker-Planck operator, we may write equation (\ref{eq:backward_fk}) as
\begin{equation}\label{eq:backward_fK_permeable}
    \fl \frac{\partial Q_r(p,t|x_0)}{\partial t}=D\frac{\partial^2 Q_r(p,t|x_0)}{\partial x_0^2}-p\Theta(x_0)Q_r(p,t|x_0)-\frac{D^2}{\kappa} \delta'(x_0)\partial_{x_0}Q_r(p,t|0),
\end{equation}
with the initial condition $Q_r(p,0|x_0)=1$. In the Laplace domain, $t\to\epsilon$, the solution to (\ref{eq:backward_fK_permeable}) is given by (see \ref{sec:fk_sol}),
\begin{equation}\label{eq:permeable_fk_solution}
    \widetilde{Q}_r(p,\epsilon|x_0)=\widetilde{\mathcal{Q}}(p,\epsilon|x_0)
    +\partial_{x_0} \widetilde{\mathcal{Q}}(p,\epsilon|0) \frac{\partial_x \widetilde{\mathcal{G}}(0,p,\epsilon|x_0) }{\frac{\kappa}{D^2}-\partial^2_{x,x_0}\widetilde{\mathcal{G}}(0,p,\epsilon|0)},
\end{equation}
where $\mathcal{G}(x,p,t|x_0)$ is the Green's function of the barrier free Feynman-Kac equation, i.e.
\begin{equation}\label{eq:brownian_fk}
    \frac{\partial \mathcal{G}(x,p,t|x_0)}{\partial t}=D\frac{\partial^2 \mathcal{G}(x,p,t|x_0)}{\partial x_0^2}-p\Theta(x_0)\mathcal{G}(x,p,t|x_0),
\end{equation}
with $\mathcal{G}(x,p,0|x_0)=\delta(x_0-x)$ and $\mathcal{Q}(p,t|x_0)=\int_{-\infty}^{\infty}\mathcal{G}(x,p,t|x_0)dx$. 

The solution of equation (\ref{eq:brownian_fk}) can be found by solving in the Laplace domain, 
\begin{equation}\label{eq:laplace_fk}
    \fl \epsilon \widetilde{\mathcal{G}}(x,p,\epsilon|x_0)-\delta(x_0-x)=\left\{
        \begin{array}{ll}
            D \frac{\partial^2}{\partial x_0^2} \widetilde{\mathcal{G}}(x,p,\epsilon|x_0)- p \widetilde{\mathcal{G}}(x,p,\epsilon|x_0) \ \ x_0>0,\\[4pt]
            D \frac{\partial^2}{\partial x_0^2} \widetilde{\mathcal{G}}(x,p,\epsilon|x_0) \ \ x_0<0.
        \end{array}
    \right .
\end{equation}
Two boundary conditions are required with equation (\ref{eq:laplace_fk}) for $x_0\to\pm\infty$. As $\mathcal{P}(t_r,t|x_0\to\infty)=\delta(t_r-t)$ and $\mathcal{P}(t_r,t|x_0\to-\infty)=\delta(t_r)$, this corresponds to $\widetilde{\mathcal{G}}(x,p,\epsilon|x_0\to \infty)=(\epsilon+p)^{-1}$ and $\widetilde{\mathcal{G}}(x,p,\epsilon|x_0\to -\infty)=\epsilon^{-1}$. In addition, we require continuity at the origin for $\widetilde{\mathcal{G}}(x,p,\epsilon|x_0)$ and its derivative \cite{das2023dynamics}, i.e. $\widetilde{\mathcal{G}}(x,p,\epsilon|0^-)=\widetilde{\mathcal{G}}(x,p,\epsilon|0^+)$ and $\lim_{x_0\to0^-}\partial_{x_0}\widetilde{\mathcal{G}}(x,p,\epsilon|x_0)=\lim_{x_0\to0^+}\partial_{x_0}\widetilde{\mathcal{G}}(x,p,\epsilon|x_0)$. The presence of the Dirac-$\delta$ function on the left-hand side (LHS) means we have two more conditions, such that we have continuity at $x_0=x$ therefore $\widetilde{\mathcal{G}}(x,p,\epsilon|x^-)=\widetilde{\mathcal{G}}(x,p,\epsilon|x^+)$ and from integrating over the Dirac-$\delta$, we have $\lim_{x_0\to x^-}\partial_{x_0}\widetilde{\mathcal{G}}(x,p,\epsilon|x_0)-\lim_{x_0\to x^+}\partial_{x_0}\widetilde{\mathcal{G}}(x,p,\epsilon|x_0)=1/D$.

Solving equation (\ref{eq:laplace_fk}) with the aforementioned conditions to obtain $\widetilde{\mathcal{G}}(x,p,\epsilon|x_0)$ and inserting into equation (\ref{eq:permeable_fk_solution}) gives $\widetilde{Q}_r(p,\epsilon|x_0)$. Setting $x_0=0^+$ and $x_0=0^-$ leads to,
\begin{equation}\label{eq:occ_plus_lap}
    \widetilde{Q}_r(p,\epsilon|0^+)=\frac{\sqrt{D} \epsilon +\kappa  \left(\sqrt{p+\epsilon }+\sqrt{\epsilon }\right)}{\sqrt{D} \epsilon  (p+\epsilon )+\kappa  \sqrt{\epsilon } \left(\sqrt{\epsilon  (p+\epsilon )}+p+\epsilon \right)},
\end{equation}
and
\begin{equation}\label{eq:occ_minus_lap}
    \widetilde{Q}_r(p,\epsilon|0^-)=\frac{\sqrt{D} (p+\epsilon )+\kappa  \left(\sqrt{p+\epsilon }+\sqrt{\epsilon }\right)}{\sqrt{D} \epsilon  (p+\epsilon )+\kappa  \sqrt{\epsilon } \left(\sqrt{\epsilon  (p+\epsilon )}+p+\epsilon \right)}.
\end{equation}
Instantly one can see that in the no barrier limit, $\kappa \to \infty$, we recover the Arcsine law, $\widetilde{Q}_r(p,\epsilon|0)=1/\sqrt{\epsilon(p+\epsilon)}$. For finite $\kappa$ the double inverse Laplace transform ($p\to t_r$ and $\epsilon \to t$) of equations (\ref{eq:occ_plus_lap}) and (\ref{eq:occ_minus_lap}) can be written in terms of the scaling relation (see 
 \ref{sec:residence_inverse_laplace}), 
\begin{equation}
    \mathcal{P}(t_r,t|0^\pm)=\frac{\kappa^2}{D}\mathdutchcal{F}_r^\pm\left(\frac{\kappa^2}{D}t_r,\frac{\kappa^2}{D}(t-t_r)\right)
\end{equation}
where,
\begin{equation}\label{eq:occ_plus_scaling}
    \mathdutchcal{F}_r^+(\tau_1,\tau_2)=\mathdutchcal{f}^+(\tau_1,\tau_2)+\mathdutchcal{g}^+(\tau_1,\tau_2)+e^{\tau_1}\mathrm{erfc}(\sqrt{\tau_1})\delta(\tau_2),
\end{equation}
and 
\begin{equation}\label{eq:occ_minus_scaling}
    \mathdutchcal{F}_r^-(\tau_1,\tau_2)=\mathdutchcal{f}^-(\tau_1,\tau_2)+\mathdutchcal{g}^-(\tau_1,\tau_2)+e^{\tau_2}\mathrm{erfc}(\sqrt{\tau_2})\delta(\tau_1).
\end{equation}
The Dirac-$\delta$ functions appear due to the particle spending the whole time in the positive region or none of the time in that region, and so are multiplied the probability of never crossing the barrier for the whole time $t$, i.e. equation (\ref{eq:prob_no_cross}). The scaling functions in equations (\ref{eq:occ_plus_scaling}) and (\ref{eq:occ_minus_scaling}) are given by (see \ref{sec:residence_inverse_laplace}),
\begin{equation}\label{eq:f_plus}
    \mathdutchcal{f}^+(\tau_1,\tau_2)=\frac{1}{\sqrt{\pi \tau_1}}e^{\tau_2}\mathrm{erfc}(\sqrt{\tau_2}),
\end{equation}
\begin{equation}\label{eq:f_minus}
    \mathdutchcal{f}^-(\tau_1,\tau_2)=\frac{2}{\pi}\sqrt{\frac{\tau_2}{\tau_1}}-2\tau_2\mathdutchcal{f}^+(\tau_1,\tau_2),
\end{equation}
and 
\begin{equation}
    \mathdutchcal{g}^\pm(\tau_1,\tau_2)=\sum_{n=0}^{\infty}\frac{(-1)^n \tau_1^{n/2}}{\Gamma(\frac{n}{2}+1)} \mathcal{C}_n^\pm(\tau_2),
\end{equation}
where $\mathcal{C}_n^+(\tau_2)$ and $\mathcal{C}_n^-(\tau_2)$ are detailed in equations (\ref{eq:c_plus}) and (\ref{eq:c_minus}) and $\Gamma(z)=\int_0^\infty t^{z-1}e^{-t}dt$ is the Gamma function.

By having the analytical expressions for $\mathcal{P}(t_r,t|0^\pm)$, we can study the asymptotics of this distribution. For the short time asymptotics, $t<<D/\kappa^2$, we take $\tau_1,\tau_2 \to 0$ in $\mathdutchcal{F}_r^\pm(\tau_1,\tau_2)$ and use $\mathdutchcal{f}^+(\tau_1,\tau_2)\simeq 1/\sqrt{\pi \tau_1}$ and $\mathdutchcal{g}^+(\tau_1,\tau_2)\simeq 2\sqrt{\tau_1/\tau_2}/\pi$ (see \ref{sec:g_asymptotics}) to give 
\begin{equation}\label{eq:short_occ_plus}
    \mathcal{P}(t_r,t|0^+)\simeq \frac{\kappa}{\sqrt{\pi D t_r}}+\frac{2\kappa^2 \sqrt{t_r}}{\pi D \sqrt{t-t_r}}+\left(1-2\kappa \sqrt{\frac{t_r}{\pi D}}\right)\delta(t-t_r).
\end{equation}
Then from equation (\ref{eq:f_minus}) we have $\mathdutchcal{f}^-(\tau_1,\tau_2)\simeq 2 \sqrt{\tau_2/\tau_1}/\pi$ and $\mathdutchcal{g}^-(\tau_1,\tau_2)\simeq 1/\sqrt{\pi \tau_2}$ (see \ref{sec:g_asymptotics}) for $\tau_1,\tau_2\to0$, which leads to the following symmetrical dependence for $t<<D/\kappa^2$, 
\begin{equation}
    \mathcal{P}(t_r,t|0^+)=\mathcal{P}(t-t_r,t|0^-).
\end{equation}

The long time asymptotics, $t>>D/\kappa^2$, corresponds to the limit $\epsilon\to 0$ in the Laplace domain, by expanding $Q_r(p,\epsilon|0^\pm)$ in equations (\ref{eq:occ_plus_lap}) and (\ref{eq:occ_minus_lap}) around small $\epsilon$ to leading order whilst keeping $p+\epsilon$ finite, one obtains $Q_r(p,\epsilon|0^\pm)\simeq 1/\sqrt{\epsilon(p+\epsilon)}$, leading to the Arcsine distribution,
\begin{equation}\label{eq:occ_arcsine}
    \mathcal{P}(t_r,t|0^\pm)\simeq\frac{1}{\pi\sqrt{t_r(t-t_r)}},
\end{equation}
showing how the influence of the permeable barrier on $\mathcal{P}(t_r,t|0^\pm)$ wanes over time. 

\begin{figure}[htp]
    \centering
    \subfloat{\includegraphics[width=0.5\textwidth]{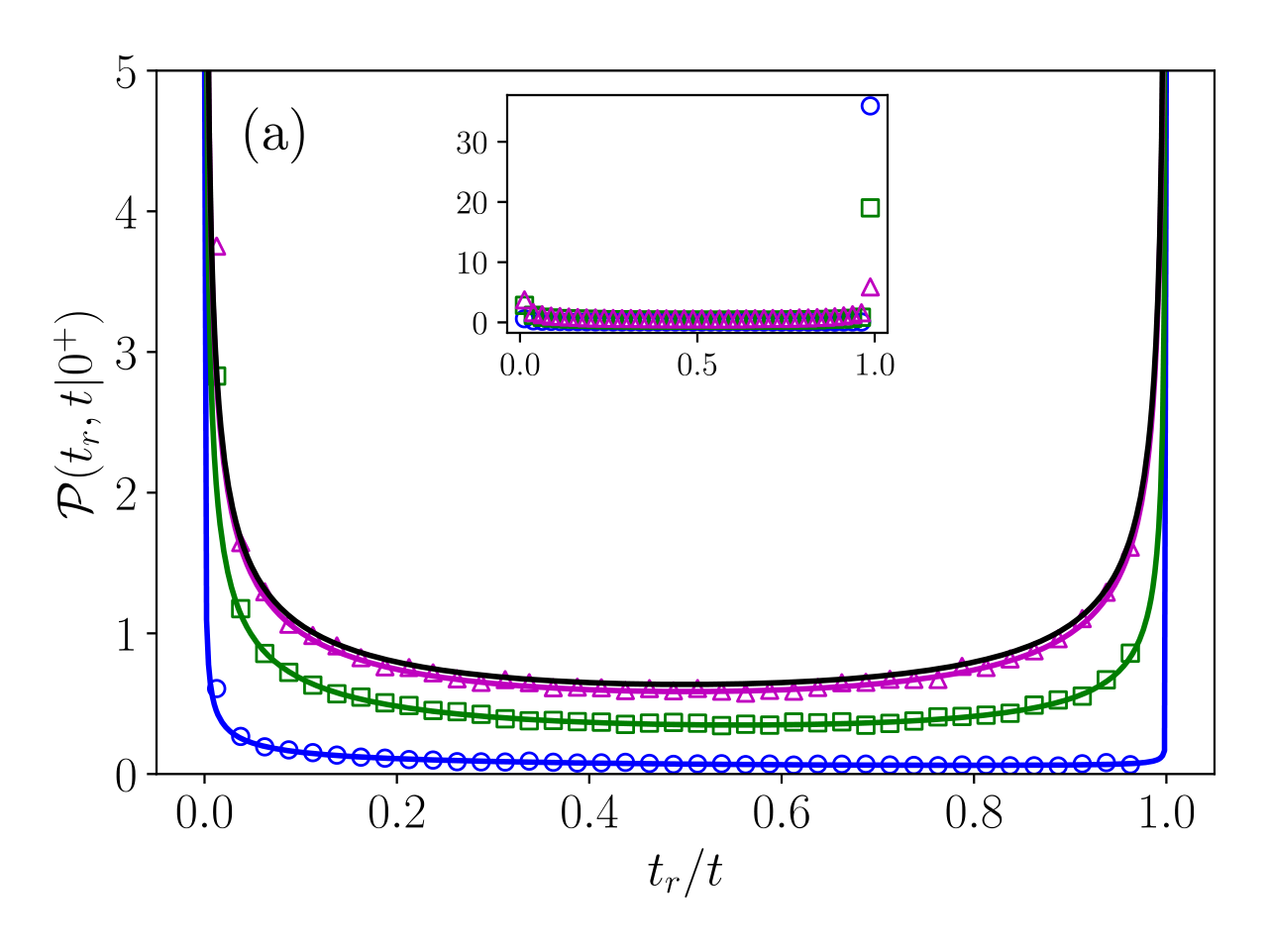}}
    \subfloat{\includegraphics[width=0.5\textwidth]{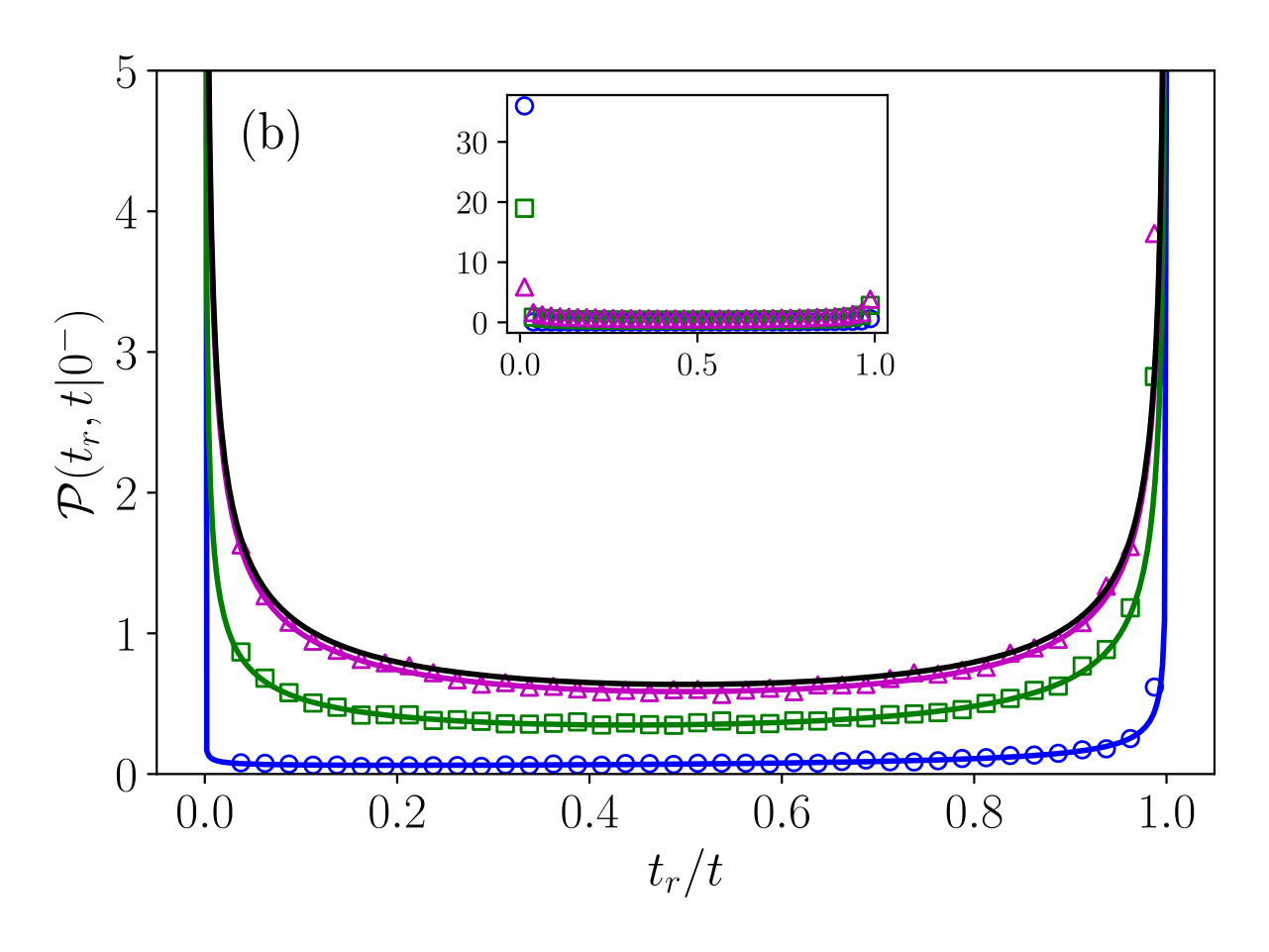}}
    \caption{Residence time density of a Brownian particle in the positive half-space, $\mathcal{P}(t_r,t|0^\pm)$, where a permeable barrier has been placed at the origin. Two distinct initial positions are considered, (a) $x_0=0^+$ and (b) $x_0=0^-$, for different values of the dimensionless parameter $\kappa^2 t/D$, all other quantities are in arbitrary units. Markers represent stochastic simulations. Once again the inset in figures (a) and (b) show the presence of the Dirac-$\delta$ from simulations (the analytic Dirac-$\delta$ is omitted from the plot). The parameter values for each curve are the same as in figure \ref{fig:max} (a).}
    \label{fig:occ}
\end{figure}

We plot $\mathcal{P}(t_r,t|0^\pm)$ in figure \ref{fig:occ} to show the excellent match with stochastic simulations. By varying the strength of the permeability of the barrier the resulting curves move further from the Arcsine distribution, where for smaller $\kappa$, the peaks at $t_r=0$ and $t_r=t$ become sharper. This illustrates the notion that once the particle crosses the barrier it is unlikely to do so again for small permeabilities. 

\section{Time of the Last Crossing of the Origin $t_\ell$}\label{sec:last_passage}

Finally, we consider the third Arcsine law, the probability density, $\mathcal{P}(t_\ell,t|0^\pm)$, of the last time the Brownian particle crosses the origin, which in our case is equivalent to passing through the permeable barrier. Unlike the previous two Arcsine laws, the probability distribution is the same for either $x_0=0^+$ or $x_0=0^-$, as the crossing process is symmetric about the origin. Thus, for simplicity we take $x_0=0^+$. To find $\mathcal{P}(t_\ell,t|0^+)$ we exploit the Markovian nature of the process and use a path decomposition approach similar to the method used in section \ref{sec:max_max_time}. Again we split the trajectory into two parts, $\{x(\tau): \tau \in [0,t_\ell]\}$ and  $\{x(\tau): \tau \in [t_\ell,t] \}$, where the first part is the trajectory that crosses the origin at $t_\ell$, and the second part is that the trajectory does not reach the origin for the remaining time $t-t_\ell$. However, the presence of a permeable barrier at the origin adds further complexity, due to the probability of reaching $0^+$ and $0^-$ being different. Therefore, we write $\mathcal{P}(t_\ell,t|0^+)$ as,
\begin{equation}\label{eq:last_passage_limit}
    \fl \mathcal{P}(t_\ell,t|0^+)=\lim_{\varepsilon\to 0^+} \frac{1}{N(\varepsilon)}\Big[P(0^-,t_\ell|0^+)S(-\varepsilon,t-t_\ell|0^-)\\
    +P(0^+,t_\ell|0^+)S(\varepsilon,t-t_\ell|0^+)\Big],
\end{equation}
where the sum in equation (\ref{eq:last_passage_limit}) comes from the last passage being an up crossing or a down crossing. We take the limit $\varepsilon\to 0^+$ to find the double Laplace transform of $\mathcal{P}(t_\ell,t|0^+)$, i.e. $\widetilde{Q}_\ell(p,\epsilon|0^+)=\int_0^\infty dt e^{-\epsilon t}\int_0^\infty dt_\ell \ e^{-pt_\ell} \mathcal{P}(t_\ell,t|0^+)$, such that
\begin{equation}\label{eq:last_passage__laplace}
    \fl \widetilde{Q}_\ell(p,\epsilon|0^+)=\lim_{\varepsilon\to 0^+} \frac{1}{N(\varepsilon)}\Big[\widetilde{P}(0^-,p+\epsilon|0^+)\widetilde{S}(-\varepsilon,\epsilon|0^-)\\
    +\widetilde{P}(0^+,p+\epsilon|0^+)\widetilde{S}(\varepsilon,\epsilon|0^+)\Big],
\end{equation}
where $\widetilde{P}(x,\epsilon|x_0)$ is given by equation (\ref{eq:diffusion_permeable_sol}) and $\widetilde{S}(\pm\varepsilon,\epsilon|0^\pm)$ is found to first order in $\varepsilon$ as previously. To find the normalization factor $N(\epsilon)$, we use the normalization condition,
\begin{equation}\label{eq:last_passage_normalization}
    \int_0^t \mathcal{P}(t_\ell,t|0^+)dt_\ell=1-S(0^-,t|0^+),
\end{equation}
which indicates that due to the presence of the permeable barrier, there is a probability that the particle will not cross the origin in a time $t$. After substituting these expressions into (\ref{eq:last_passage__laplace}) and using (\ref{eq:last_passage_normalization}) we find $N(\varepsilon)=\varepsilon/D$, giving
\begin{equation}
    \widetilde{Q}_\ell(p,\epsilon|0^+)=\frac{\kappa}{\sqrt{\epsilon(p+\epsilon)}(\sqrt{D \epsilon}+\kappa)},
\end{equation} 
using standard inverse Laplace transform relations \cite{roberts1966table}, we obtain 
\begin{equation}\label{eq:last_passage_scale}
    \mathcal{P}(t_\ell,t|0^+)=\frac{\kappa^2}{D}\mathdutchcal{F}_\ell\left(\frac{\kappa^2}{D}t_\ell,\frac{\kappa^2}{D}(t-t_\ell)\right),
\end{equation}
where 
\begin{equation}
    \mathdutchcal{F}_\ell(\tau_1,\tau_2)=\mathdutchcal{f}^+(\tau_1,\tau_2)
\end{equation}
for $\mathdutchcal{f}^+(\tau_1,\tau_2)$ defined in equation (\ref{eq:f_plus}). The striking feature here is that the scaling function that fully describes $\mathcal{P}(t_\ell,t|0^+)$ is also part of the scaling function describing the residence time density, $\mathcal{P}(t_r,t|0^+)$. Comparing the short time limit, $t<<D/\kappa^2$,
\begin{equation}
    \mathcal{P}(t_\ell,t|0^+)\simeq \frac{\kappa}{\sqrt{\pi D t_\ell}},
\end{equation}
and the long time limit, $t>>D/\kappa^2$ for $t_\ell<<t$ , 
\begin{equation}\label{eq:last_passage_long}
    \mathcal{P}(t_\ell,t|0^+)\simeq \frac{1}{\pi\sqrt{t_\ell(t-t_\ell)}},
\end{equation}
to that of $\mathcal{P}(t_r,t|0^+)$, i.e. equations (\ref{eq:short_occ_plus}) and (\ref{eq:occ_arcsine}), we see that the peaks at the origin of both distributions have the same time dependence. This feature can be understood by considering that for instantaneous last crossing times, $t_\ell << t$, this crossing event is almost certainly going to be the first and last crossing, which corresponds to the residence time being equivalent to the last crossing time $t_r=t_\ell$. This breaks down as $t_\ell$ gets larger, causing the different dependencies of the distributions. In this case equation (\ref{eq:last_passage_long}) is not valid for $t-t_\ell \to 0$ as $\mathdutchcal{f}^+(\tau_1,0)=1/\sqrt{\pi \tau_1}$.

\begin{figure}[htp]
    \includegraphics[width=0.5\textwidth]{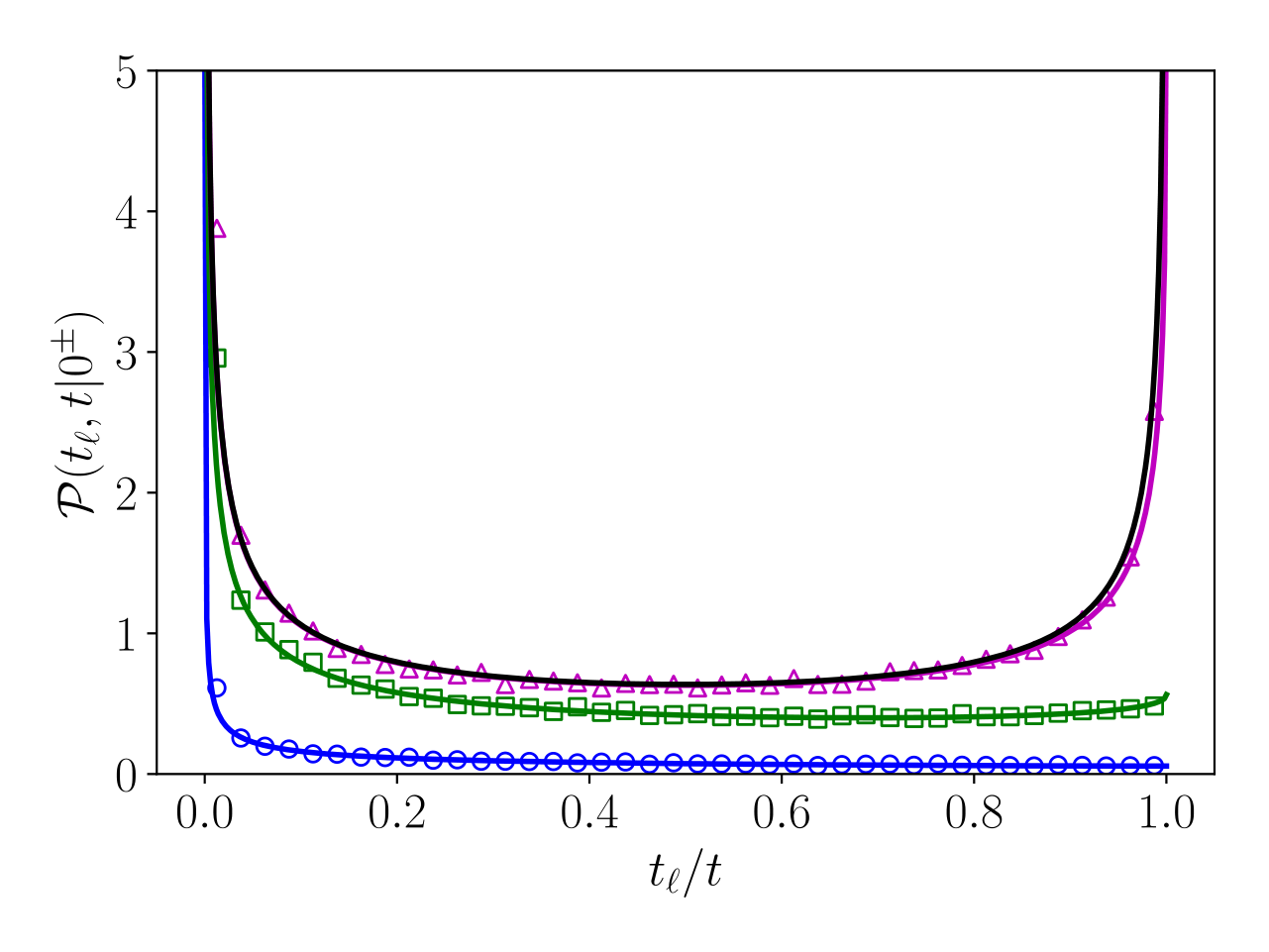}
    \centering
    \caption{Last-passage time density of a Brownian particle, $\mathcal{P}(t_\ell,t|0^\pm)$, starting to the left or right of a permeable barrier placed at the origin. The curves represent different values of the dimensionless parameter, $\kappa^2t/D$, while all other quantities are in arbitrary units. Markers represent stochastic simulations. The parameter values for each curve are the same as in figure \ref{fig:max} (a).}
    \label{fig:last_passage}
\end{figure}

We plot equation (\ref{eq:last_passage_scale}) in figure \ref{fig:last_passage} to compare with stochastic simulations, where we see an excellent match. In comparing different $\kappa$ values the main feature is the presence of a sharper peak at $t_\ell=0$ the smaller the permeability, indicating that once the particle crosses the barrier it is unlikely to do so again. Differently from the Arcsine distribution (\ref{eq:arcsine_law}), there is no divergence at $t_\ell=t$, which is due to there being a non-zero probability of not crossing the barrier for every interaction. We note that instead if one is interested in the distribution of the particle returning to the same side of the barrier for the last time, one would recover the Arcsine distribution. 

\section{Conclusion}\label{sec:conclusion}
In summary, we have investigated the extreme-value statistics and Arcsine laws of Brownian motion in the presence of a permeable barrier at the origin, using an inhomogeneous diffusion equation which accounts for the presence of a permeable barrier. The presence of this barrier requires considering two initial positions, the right and left-hand side of the barrier, i.e. $x_0=0^+$ and $x_0=0^-$, respectively. 

Firstly, using a path-decomposition technique we have obtained the joint density of the maximum displacement, $M(t)$, and the time to reach it, $t_m(t)$, for the two initial conditions, $\mathcal{P}(M,t_m,t|0^\pm)$. We have found that this quantity can be represented by the multiplication of two different scaling functions, indicating the distribution is no longer symmetric under a $t_m\to t-t_m$ transformation. For the $x_0=0^-$ case a Dirac-$\delta$ centered at $t_m=0$ is also present due to the probability of not passing through the barrier in finite time. At short times, $t<<D/\kappa^2$, $\mathcal{P}(M,t_m,t|0^+)$ can be asymptotically approximated by the distribution for when the barrier is fully reflecting, $\kappa=0$, which is given in terms of Jacobi Theta functions. This approximation is valid due to a very small number of particles mananging to go through the barrier for short times. For $\mathcal{P}(M,t_m,t|0^-)$ we find a stronger approximation than just the reflecting barrier distribution, which is $\delta(M)\delta(t_m)$, which accounts for the very few particles that do cross the barrier. 

We have also investigated the respective marginal distributions, $\mathcal{P}(M,t|0^\pm)$ and $\mathcal{P}(t_m,t|0^\pm)$. The presence of the barrier has a large impact on the monotonicity of the distribution, $\mathcal{P}(M,t|0^\pm)$, where for certain permeabilities a maximum appears away from the origin. At long times, $t>>D/\kappa^2$, the distribution is still dependent on $\kappa$. For $\mathcal{P}(t_m,t|0^\pm)$ the distribution remains asymmetric for long times, such that for large $t_m$ the usual Arcsine distribution is recovered, whereas we get a different dependence for $t_m\to 0$. 

For the rest of the paper we have investigated the other two Arcsine laws, namely the distributions of the residence time, $t_r(t)$, and the last crossing of the origin, $t_\ell(t)$. Using Feynman-Kac theory we have calculated $\mathcal{P}(t_r,t|0^\pm)$ analytically and have found the dependence in terms of a scaling function. For $x_0=0^+$ and $x_0=0^-$ we have a Dirac-$\delta$ located at $t_r=t$ and $t_r=0$, respectively, due to the particle never crossing the barrier. In the short time limit we have found that the scaling functions for $x_0=0^\pm$ are equivalent under the transformation $t_r\to t-t_r$, and in the long time limit we recover the Arcsine distribution. 

Finally, we have studied $\mathcal{P}(t_\ell,t|0^\pm)$, which is equivalent for either initial condition $x_0=0^+$ or $x_0=0^-$, since $t_\ell(t)$ is a crossing event. Taking into account up and down crossings we have found $\mathcal{P}(t_\ell,t|0^\pm)$ in terms of known functions. Interestingly the scaling function that describes this distribution happens to be part of the scaling function which describes $\mathcal{P}(t_r,t|0^+)$, where the peak at the origin of both distributions have the same dependence, because for $t_\ell<<t$ the first crossing very likely corresponds to the last crossing, which implies $t_\ell=t_r$. As $t_\ell$ becomes larger this is no longer the case, and we do not observe a divergence as $t_\ell \to t$, due to the non-zero probability of not crossing the barrier for any interaction. 

Possible extensions of this study would include the analysis of how a permeable barrier affects the time, $T$, between the maximum and minimum of the process \cite{mori2019time,mori2020distribution}. It would be interesting to see if the presence of a barrier breaks the symmetry around $T=0$ and whether the initial position i.e. $x_0=0^\pm$ leads to differences in the respective distributions. Another interesting avenue to explore is changing the underlying Brownian motion to a different stochastic process, such as anomalous subdiffusion (where an equation akin to (\ref{eq:diffusion_permeable}) has been found for this case \cite{kay2023subdiffusion}), and what is the impact of a permeable barrier to the unperturbed statistics. Futhermore, in a recent study the joint distribution of the first passage time to a target and the number of distinct sites visited when the target is reached of a random walk has been obtained \cite{klinger2022joint}. It would be of interest to quantify how the presence of a permeable barrier affects this distribution, where specifically one would be able to glean the impact of a spatial heterogeneity on the kinetics and geometry of space exploration.

\ack
TK and LG acknowledge funding from, respectively, an Engineering and Physical Sciences Research Council (EPSRC) DTP student grant and the Biotechnology and Biological Sciences Research Council (BBSRC) Grant No. BB/T012196/1 and NERC Grant No. NE/W00545X/1. This work was carried out using the computational facilities of the Advanced Computing Research Centre, University of Bristol - http://www.bristol.ac.uk/acrc/.
\\

\appendix

\section{Solution of the Feynman-Kac Equation}\label{sec:fk_sol}

Here we show how the solution of equation (\ref{eq:backward_fK_permeable}), $Q_r(p,t|x_0)$, can be represented in terms the Green's function, $\mathcal{G}(x,p,t|x_0)$, of the barrier free Feynman-Kac equation, (\ref{eq:brownian_fk}). If we take the last term on the right-hand side (RHS) of equation (\ref{eq:backward_fK_permeable}) as an inhomogeneous term, we may construct the solution as follows,
\begin{eqnarray}
    Q_r(p,t|x_0)&=\int_{-\infty}^{\infty} dy \mathcal{G}(y,p,t|x_0) Q_r(p,0|y)\\ \nonumber 
    &- \frac{D^2}{\kappa} \int_0^t dt' \int_{-\infty}^{\infty} dy \mathcal{G}(y,p,t-t'|x_0) \delta'(x_0) \partial_{x_0}Q_r(p,t'|0).
\end{eqnarray}
Using $Q_r(p,0|y)=1$ with $\mathcal{Q}(p,t|x_0)=\int_{-\infty}^\infty \mathcal{G}(y,p,t|x_0)$ and Laplace transforming, $t\to \epsilon$, we have
\begin{equation}\label{eq:fk_laplace_greens}
    \widetilde{Q}_r(p,\epsilon|x_0)=\widetilde{\mathcal{Q}}(p,\epsilon|x_0)+\frac{D^2}{\kappa} \partial_x \widetilde{\mathcal{G}}(0,p,\epsilon|x_0) \partial_{x_0} \widetilde{Q}_r(0,\epsilon|0).
\end{equation}
Then by taking the derivative of both sides of equation (\ref{eq:fk_laplace_greens}) with respect to $x_0$ and setting $x_0=0$, we find
\begin{equation}\label{eq:fk_sol_deriv}
    \partial_{x_0} \widetilde{Q}_r(0,\epsilon|0)=\frac{\partial_{x_0}\widetilde{\mathcal{Q}}(p,\epsilon|0)}{1-\frac{D^2}{\kappa}\partial^2_{x,x_0}\widetilde{\mathcal{G}}(0,p,\epsilon|0)},
\end{equation}
then after inserting equation (\ref{eq:fk_sol_deriv}) into (\ref{eq:fk_laplace_greens}) we obtain (\ref{eq:permeable_fk_solution}).

\section{Double Inverse Laplace Transform of $\widetilde{Q}_m(M,p,\epsilon|0^\pm)$}\label{sec:joint_max_inverse_laplace}

To perform the Laplace inversion of $\widetilde{Q}_m(M,p,\epsilon|0^+)$, we write equation (\ref{eq:max_plus_double_laplace}) as 
\begin{equation}
    \widetilde{Q}_m(M,p,\epsilon|0^+)=\frac{1}{\kappa}\widetilde{\mathdutchcal{G}}^+\left(\frac{\kappa}{D}M,\frac{D}{\kappa^2}(p+\epsilon)\right)\widetilde{\mathdutchcal{H}}\left(\frac{\kappa}{D}M,\frac{D}{\kappa^2}\epsilon\right),
\end{equation}
then $\mathcal{P}(M,t_m,t|0^+)$ is given by the two Bromwich integrals,
\begin{eqnarray}\label{eq:max_joint_bromwich}
    \mathcal{P}(M,t_m,t|0^+)&=\frac{\kappa^3}{D^2} \frac{1}{(2\pi i)^2} \int_{\gamma_1 -i \infty}^{\gamma_1+i\infty} ds_1 e^{\frac{\kappa^2}{D}t_m s_1 } \widetilde{\mathdutchcal{G}}^+\left(\frac{\kappa}{D}M,s_1\right)  \\\nonumber 
    & \times \int_{\gamma_2 -i \infty}^{\gamma_2+i\infty} ds_2 e^{\frac{\kappa^2}{D}(t-t_m) s_2} \widetilde{\mathdutchcal{H}}\left(\frac{\kappa}{D}M,s_2\right),
\end{eqnarray}
where $\gamma_1$ and $\gamma_2$ are greater than the real part of all singularities of $\widetilde{\mathdutchcal{G}}^+\left(\frac{\kappa}{D}M,s_1\right)$ and $\widetilde{\mathdutchcal{H}}\left(\frac{\kappa}{D}M,s_2\right)$, respectively and $\widetilde{\mathdutchcal{G}}^+(y,s_1)$ and $\widetilde{\mathdutchcal{H}}(y,s_2)$ are given by,
\begin{equation}\label{eq:g_plus_dimensionless}
    \widetilde{\mathdutchcal{G}}^+(y,s_1)=\frac{e^{y\sqrt{s_1}}(\sqrt{s_1}+1)}{\sqrt{s_1}+e^{2y\sqrt{s_1}}(\sqrt{s_1}+2)}
\end{equation}
and 
\begin{equation}\label{eq:h_dimensionless}
    \widetilde{\mathdutchcal{H}}(y,s_2)=\frac{2\left(-\sqrt{s_2}+e^{2y\sqrt{s_2}}(\sqrt{s_2}+2)\right)}{\sqrt{s_2}\left(\sqrt{s_2}+e^{2 y\sqrt{s_2}}(\sqrt{s_2}+2)\right)}.
\end{equation}
To find $\mathcal{P}(M,t_m,t|0^+)$ we require the following Laplace inversions $\mathcal{L}^{-1}_{s_1\to\tau_1}\left\{ \widetilde{\mathdutchcal{G}}^+(y,s_1) \right\}$ and $\mathcal{L}^{-1}_{s_2\to\tau_2}\left\{ \widetilde{\mathdutchcal{H}}(y,s_2) \right\}$, then $\mathcal{P}(M,t_m,t|0^-)$ only requires the Laplace inversion, $\mathcal{L}^{-1}_{s_1\to\tau_1}\left\{ \widetilde{\mathdutchcal{G}}^-(y,s_1) \right\}$, where $\widetilde{\mathdutchcal{G}}^-(y,s_1)=(\sqrt{s_1}+1)^{-1}\widetilde{\mathdutchcal{G}}^+(y,s_1)$.

\subsection{Laplace Inversion of $\widetilde{\mathdutchcal{G}}^\pm(y,s_1)$}

From the definition of $\widetilde{\mathdutchcal{G}}^+(y,s_1)$ in equation (\ref{eq:g_plus_dimensionless}) we see that $\widetilde{\mathdutchcal{G}}^+(y,s_1)$ has no poles but has a branch point at $s_1=0$. Thus, by taking the branch cut to be the negative real axis, then $\widetilde{\mathdutchcal{G}}^+(y,s_1)$ is analytic inside the contour, $C$, in figure \ref{fig:contour}, meaning that $\oint_{C} e^{s_1 \tau_1} \widetilde{\mathdutchcal{G}}^+(y,s_1) ds_1=0$. Then by taking $R\to \infty$ and $\alpha\to 0$ for the contour $C$ the Laplace inversion is given by,
\begin{equation}\label{eq:g_plus_contour}
    \fl\mathcal{L}^{-1}_{s_1\to \tau_1}\left\{\widetilde{\mathdutchcal{G}}^+(y,s_1)\right\}=\frac{1}{2\pi i} \int_{C_1} e^{s_1 \tau_1} \widetilde{\mathdutchcal{G}}^+(y,s_1) ds =\frac{-1}{2\pi i}\left(\int_{C_3}+\int_{C_5}\right) e^{s_1 \tau_1} \widetilde{\mathdutchcal{G}}^+(y,s_1) ds_1
\end{equation}
because in the limit $R\to\infty$ the contributions from $C_2$ and $C_6$ vanish, and for $\alpha\to 0$ the contribution from $C_4$ is zero. Therefore, all we require is finding the integrals $\int_{C_3}$ and $\int_{C_5}$. 

For $\int_{C_3}$ we let $s_1=ze^{i \pi}$, giving
\begin{equation}
    \int_{C_3} e^{s_1 \tau_1} \widetilde{\mathdutchcal{G}}^+(y,s_1) ds_1= -\int_{R}^{\alpha} e^{- \tau_1 z }  \frac{e^{i y\sqrt{z}}(i\sqrt{z}+1)}{i\sqrt{z}+e^{2iy\sqrt{z}}(i\sqrt{z}+2)}dz,
\end{equation}
and for $\int_{C_5}$ we let $s_1=ze^{-i \pi}$, leading to
\begin{equation}
    \int_{C_5} e^{s_1 \tau_1} \widetilde{\mathdutchcal{G}}^+(y,s_1) ds_1= -\int_{\alpha}^{R} e^{- \tau_1 z }  \frac{e^{-i y\sqrt{z}}(-i\sqrt{z}+1)}{-i\sqrt{z}+e^{-2iy\sqrt{z}}(-i\sqrt{z}+2)}dz.
\end{equation}
Taking $R\to\infty$ and $\alpha\to 0$, substituting back into equation (\ref{eq:g_plus_contour}) and converting the complex exponentials to trigonometric functions, we obtain $\mathdutchcal{G}^+(y,\tau_1)$ as defined in equation (\ref{eq:g_plus}).

Similarly, since $\widetilde{\mathdutchcal{G}}^-(y,s_1)=(\sqrt{s_1}+1)^{-1}\widetilde{\mathdutchcal{G}}^+(y,s_1)$, one can see that by altering the above calculations it leads to the definition of $\mathdutchcal{G}^-(y,\tau_1)$ in equation (\ref{eq:g_minus}).

\subsection{Laplace Inversion of $\widetilde{\mathdutchcal{H}}(y,s_2)$}

\begin{figure}
    \includegraphics[width=0.5\textwidth]{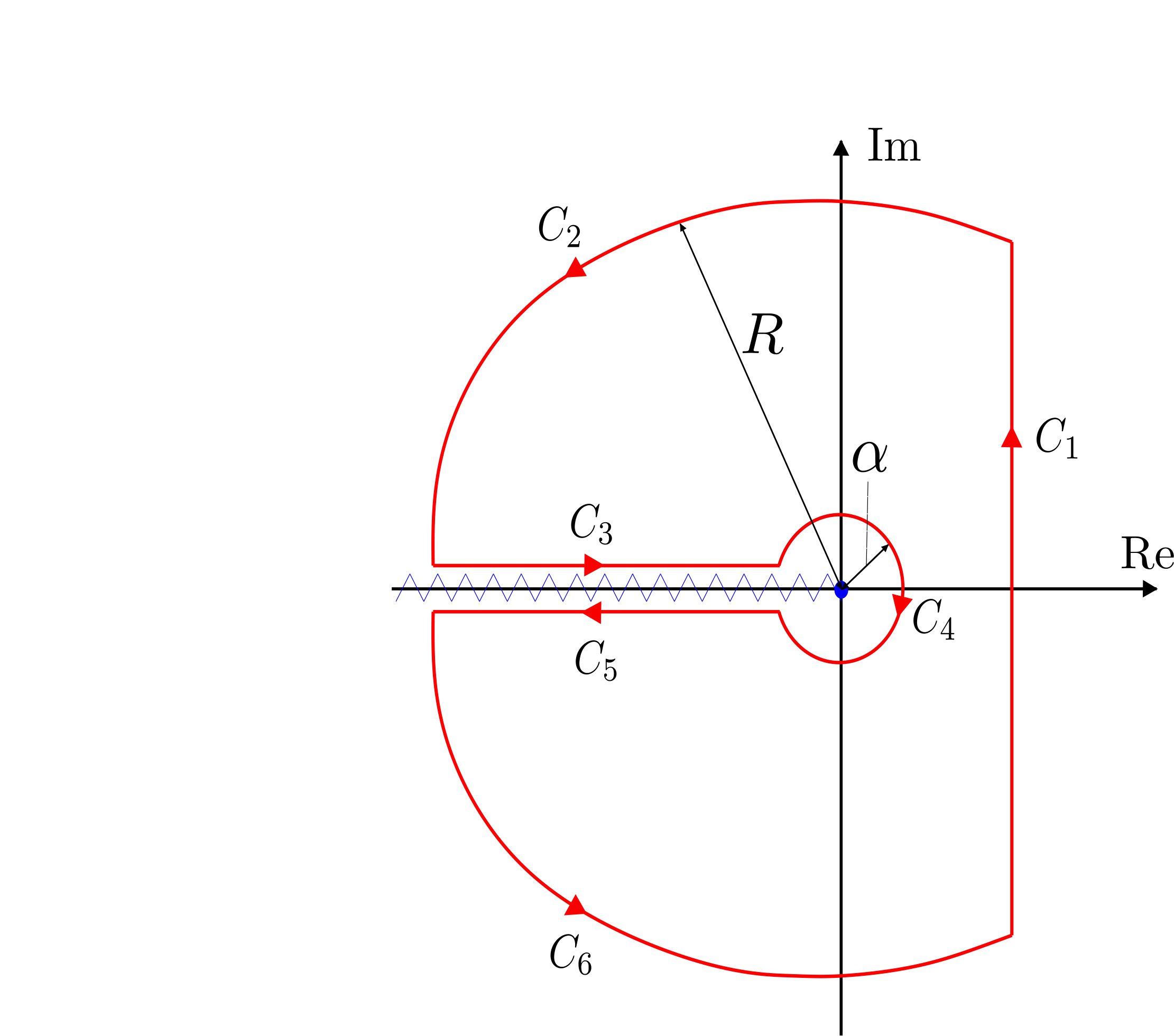}
    \centering
    \caption{Key-hole contour, $C=C_1+...+C_6$, used for the Bromwich integration to perform the inverse Laplace transforms of various functions defined in equations (\ref{eq:g_plus_contour}), (\ref{eq:h_contour}), (\ref{eq:I_plus_contour}) and (\ref{eq:I_minus_contour}), where the only singularity is a branch point at the origin.}
    \label{fig:contour}
\end{figure}

From equation (\ref{eq:h_dimensionless}) one can see that $\widetilde{\mathdutchcal{H}}(y,s_2)$ has a branch point at $s_2=0$. Proceeding similarly to the above case, we use the contour in figure \ref{fig:contour} and obtain
\begin{equation}\label{eq:h_contour}
    \fl\mathcal{L}^{-1}_{s_2\to \tau_2}\left\{\widetilde{\mathdutchcal{H}}(y,s_2)\right\}=\frac{1}{2\pi i} \int_{C_1} e^{s \tau_2} \widetilde{\mathdutchcal{H}}(y,s_2) ds =\frac{-1}{2\pi i}\left(\int_{C_3}+\int_{C_5}\right) e^{s_2 \tau_2} \widetilde{\mathdutchcal{H}}(y,s_2) ds_2,
\end{equation}
because in the limit $R\to\infty$ and $\alpha\to0$ the contributions from $C_2$, $C_6$ and $C_4$ vanish, and we have
\begin{equation}
    \int_{C_3} e^{s_2 \tau_2} \widetilde{\mathdutchcal{H}}(y,s_2) ds_2= -\int_{R}^{\alpha} e^{- \tau_2 z }  \frac{2\left(-i\sqrt{z}+e^{2i\sqrt{z}}(i\sqrt{z}+2)\right)}{i\sqrt{z}\left(i\sqrt{z}+e^{2 i\sqrt{z}}(i\sqrt{z}+2)\right)}dz,
\end{equation}
and
\begin{equation}
    \fl \int_{C_5} e^{s_2 \tau_2} \widetilde{\mathdutchcal{H}}(y,s_2) ds_2= -\int_{\alpha}^{R} e^{- \tau_2 z }  \frac{2\left(i\sqrt{z}+e^{-2i\sqrt{z}}(-i\sqrt{z}+2)\right)}{-i\sqrt{z}\left(-i\sqrt{z}+e^{-2 i\sqrt{z}}(-i\sqrt{z}+2)\right)}dz.
\end{equation}
After taking $R\to\infty$, $\alpha\to0$ and substituting these expressions into equation (\ref{eq:h_contour}) one can see that we recover $\mathdutchcal{H}(y,\tau_2)$ in equation (\ref{eq:max_H}).

\section{Inverse Laplace Transform of $\widetilde{Q}_m(M,\epsilon|0^\pm)$}\label{sec:max_inverse_laplace}

The marginal over $t_m$ of $\mathcal{P}(M,t_m,t|0^+)$ corresponds to $\widetilde{Q}_m(M,0,\epsilon|0^+)$ and from equations (\ref{eq:max_joint_bromwich}), (\ref{eq:g_plus_dimensionless}) and (\ref{eq:h_dimensionless}), we are looking for the Laplace inversion, $\mathcal{L}^{-1}_{s \to \tau} \left\{ \widetilde{\mathcal{I}}^+(y,s)\right\}$, where 
\begin{equation}
    \widetilde{\mathcal{I}}^+(y,s)=\frac{2e^{y\sqrt{s}}(\sqrt{s}+1)(-\sqrt{s}+e^{2y\sqrt{s}}(\sqrt{s}+2))}{\sqrt{s}(\sqrt{s}+e^{2y\sqrt{s}}(\sqrt{s}+2))^2}.
\end{equation}
Once again we have a branch point at the origin and thus use the contour $C$ in figure \ref{fig:contour} to perform the Laplace inversion. The contributions from $C_2$, $C_6$ and $C_4$ vanish, leaving us to calculate $\int_{C_3}$ and $\int_{C_5}$. By using the substitutions, $s=ze^{i\pi}$ and $s=ze^{-i\pi}$ for $C_3$ and $C_5$ respectively, whilst taking $R\to \infty$ and $\alpha\to 0$, we have
\begin{equation}\label{eq:I_plus_contour}
    \fl\mathcal{L}^{-1}_{s\to\tau}\left\{\widetilde{\mathcal{I}}^+(y,s) \right\}=\frac{-1}{2\pi i}\left(\int_{C_3}+\int_{C_5}\right) e^{s \tau} \widetilde{\mathcal{I}}^+(y,s) ds =\frac{1}{\pi}\int_0^\infty \frac{e^{-\tau z}}{\sqrt{z}}j^+(y,z)h^2(y,z)dz,
\end{equation}
where 
\begin{equation}\label{eq:j_plus}
    j^+(y,z)=(5z+4)\cos(y\sqrt{z})-z\cos(3y\sqrt{z})+8\sqrt{z}\sin^3(y\sqrt{z}).
\end{equation}

Similarly, since $\widetilde{\mathcal{I}}^-(y,s)=(\sqrt{s}+1)^{-1}\widetilde{\mathcal{I}}^+(y,s)+(\sqrt{s}+s)^{-1}\delta(y)$, we find
\begin{eqnarray}\label{eq:I_minus_contour}
    \mathcal{L}^{-1}_{s\to\tau}\left\{\widetilde{\mathcal{I}}^-(y,s) \right\}&=\frac{-1}{2\pi i}\left(\int_{C_3}+\int_{C_5}\right) e^{s \tau} \widetilde{\mathcal{I}}^-(y,s) ds \\ \nonumber 
    &=\frac{1}{\pi}\int_0^\infty \frac{e^{-\tau z}}{\sqrt{z}}j^-(y,z)h^2(y,z)dz + e^\tau\mathrm{erfc}(\tau)\delta(y),
\end{eqnarray}
where,
\begin{equation}\label{eq:j_minus}
    \fl j^-(y,z)=(4-z)\cos(y\sqrt{z})-3z\cos(3y\sqrt{z})+2\sqrt{z}\left[z+(z-2)\cos(2y\sqrt{z})\right]\sin(y\sqrt{z}).
\end{equation}

\section{Double Inverse Laplace Transform of $\widetilde{Q}_r(p,\epsilon|0^\pm)$}\label{sec:residence_inverse_laplace}

Starting from equations (\ref{eq:occ_plus_lap}) and (\ref{eq:occ_minus_lap}), we may write
\begin{equation}
    \widetilde{Q}_r(p,\epsilon|0^\pm)=\frac{D}{\kappa^2}\widetilde{\mathdutchcal{F}}_r^\pm\left(\frac{D}{\kappa^2}(p+\epsilon),\frac{D}{\kappa^2}\epsilon\right),
\end{equation}
therefore 
\begin{equation}
    \mathcal{P}(t_r,t|0^\pm)=\frac{\kappa^2}{D}\mathcal{L}^{-1}_{s_2\to \tau_2}\left\{\mathcal{L}^{-1}_{s_1\to \tau_1}\left\{\widetilde{\mathdutchcal{F}}_r^\pm(s_1,s_2) \right\} \right\},
\end{equation}
for $\tau_1=\kappa^2 t_r/D$ and $\tau_2=\kappa^2 (t-t_r)/D$, where
\begin{equation}\label{eq:f_plus_laplace}
    \widetilde{\mathdutchcal{F}}_r^+(s_1,s_2)=\frac{s_2+\sqrt{s_1}+\sqrt{s_2}}{s_1 s_2+\sqrt{s_2} \left(\sqrt{s_1 s_2}+s_1\right)}
\end{equation}
and
\begin{equation}\label{eq:f_minus_laplace}
    \widetilde{\mathdutchcal{F}}_r^-(s_1,s_2)=\frac{s_1+\sqrt{s_1}+\sqrt{s_2}}{s_1 s_2+\sqrt{s_2} \left(\sqrt{s_1 s_2}+s_1\right)}.
\end{equation}
We now proceed to compute the Laplace inversions, $s_1\to \tau_1$ and $s_2\to\tau_2$, of equations (\ref{eq:f_plus_laplace}) and (\ref{eq:f_minus_laplace}).

\subsection{Double Laplace Inversion of $\widetilde{\mathdutchcal{F}}_r^+(s_1,s_2)$}

Let us first write equation (\ref{eq:f_plus_laplace}) as 
\begin{equation}
    \widetilde{\mathdutchcal{F}}_r^+(s_1,s_2)=\frac{1}{\sqrt{s_1} \left(s_2+\sqrt{s_2}\right)+s_2}+\frac{s_2+\sqrt{s_2}}{s_1 \left(s_2+\sqrt{s_2}\right)+\sqrt{s_1} s_2},
\end{equation}
and using the following inverse Laplace transform relations \cite{roberts1966table}: 
\begin{equation}\label{eq:laplace_inv_erfc1}
    \mathcal{L}^{-1}_{s_1\to\tau_1}\left\{\frac{1}{\sqrt{s_1}+a}\right\}=\frac{1}{\sqrt{\pi \tau_1 } }-a e^{a^2 \tau_1 } \mathrm{erfc}\left(a \sqrt{\tau_1 },\right)
\end{equation}
and
\begin{equation}\label{eq:laplace_inv_erfc2}
    \mathcal{L}^{-1}_{s_1\to\tau_1}\left\{\frac{1}{\sqrt{s_1}(\sqrt{s_1}+a)}\right\}=e^{a^2 \tau_1 } \mathrm{erfc}\left(a \sqrt{\tau_1 }\right),
\end{equation}
we obtain
\begin{eqnarray}\label{eq:f_plus_mixed}
    \mathcal{L}^{-1}_{s_1\to\tau_1}\left\{\widetilde{\mathdutchcal{F}}_r^+(s_1,s_2)\right\}&=\frac{1}{\sqrt{\pi \tau_1} \left(s_2+\sqrt{s_2}\right) } + \left(1-\frac{1}{\left(\sqrt{s_2}+1\right)^2}\right)\\ \nonumber
    &\times e^{\frac{s_2 \tau_1}{\left(\sqrt{s_2}+1\right)^2}} \mathrm{erfc}\left(\frac{\sqrt{s_2 \tau_1}}{\sqrt{s_2}+1}\right).
\end{eqnarray}
Taking the inverse Laplace transform of the first term on the left-hand side (LHS) of equation (\ref{eq:f_plus_mixed}) gives equation (\ref{eq:f_plus}). To take the inverse Laplace transform of the second term we use the following series representation,
\begin{equation}\label{eq:erfc_taylor}
    e^{z^2}\mathrm{erfc}(z)=\sum_{n=0}^{\infty} \frac{(-1)^n z^n}{\Gamma\left(\frac{n}{2}+1\right)},
\end{equation}
where we then have
\begin{eqnarray}\label{eq:erfc_series_representation}
    \fl \left(1-\frac{1}{\left(\sqrt{s_2}+1\right)^2}\right)e^{\frac{s_2 \tau_1}{\left(\sqrt{s_2}+1\right)^2}} \mathrm{erfc}\left(\frac{\sqrt{s_2 \tau_1}}{\sqrt{s_2}+1}\right)&=\left(1+\frac{\widetilde{\zeta}_1(\sqrt{s_2})}{\sqrt{s_2}}\right) \\ \nonumber
    &\times \sum_{n=0}^{\infty} \frac{(-1)^n\tau_1^{n/2}}{\Gamma\left(\frac{n}{2}+1\right)} \widetilde{\zeta}_{n+1}(\sqrt{s_2}),
\end{eqnarray}
where 
\begin{equation}
    \widetilde{\zeta}_n(\sqrt{s_2})=\left(\frac{\sqrt{s_2}}{1+\sqrt{s_2}}\right)^n.
\end{equation}
We now proceed to calculate the inversion, $\mathcal{L}^{-1}_{s_2\to\tau_2}\left\{\widetilde{\zeta}_n(\sqrt{s_2})\right\}$, where we first find the inversion of $\mathcal{L}^{-1}_{s_2\to\tau_2}\left\{\widetilde{\zeta}_n(s_2)\right\}$. We do this by utilizing the following property,
\begin{equation}\label{eq:laplace_deriv}
    \mathcal{L}_{\tau \to s}\left\{\frac{d^n}{d \tau^n} f(\tau)\right\}= s^n \widetilde{f}(s)-\sum_{m=1}^n s^{n-m} \lim_{\tau\to0} \frac{d^{m-1}}{d \tau^{m-1}}f(\tau),
\end{equation}
for some arbitrary function $f(\tau)$. Using $\mathcal{L}^{-1}_{s\to \tau}\{(1+s)^{-n} \}=\tau^{n-1}e^{-\tau}/\Gamma(n)$, and from the generalized product rule we have 
\begin{equation}
    \frac{d^n}{d \tau^n}\frac{\tau^{n-1}e^{-\tau}}{\Gamma(n)}=e^{-\tau}\sum_{k=0}^n (-1)^k {n\choose k} \frac{\tau^{k-1}}{\Gamma(k)}=-n {}_1F_1(n+1,2,-\tau),
\end{equation}
where ${}_1F_1(a,b,z)$ is the Kummer confluent hypergeometric function \cite{bateman1953higher1}. From, $\frac{d^{m-1}}{d \tau^{m-1}}\tau^{n-1}e^{-\tau}/\Gamma(n)=\tau^{n-m}{}_1F_1(n,n+m-1,-\tau)$, and since for $\tau\to0$ this is only non-zero for $m=n$, we have,
\begin{equation}
    \mathcal{L}^{-1}_{s_2\to\tau_2}\left\{\widetilde{\zeta}_n(s_2)\right\}=-n{}_1F_1(n+1,2,-\tau_2)+\delta(\tau_2).
\end{equation} 

To find $\mathcal{L}^{-1}_{s_2\to\tau_2}\left\{\widetilde{\zeta}_n(\sqrt{s_2})\right\}$ we use the property \cite{roberts1966table}:
\begin{equation}\label{eq:laplace_square_root}
    \mathcal{L}^{-1}_{s\to\tau}\left\{ \widetilde{f}(\sqrt{s})\right\}=\frac{1}{\sqrt{4\pi \tau^3}}\int_0^\infty u e^{-\frac{u^2}{4\tau}} f(u)du,
\end{equation}
and calculate the following integral as 
\begin{eqnarray}
    \fl \frac{1}{\sqrt{4 \pi  \tau ^3}}\int_0^{\infty } u e^{-\frac{u^2}{4 \tau }} \, _1F_1(n+1;2;-u) \, du&=\frac{1}{\sqrt{\pi \tau }}\, _2F_2\left(\frac{n}{2}+\frac{1}{2},\frac{n}{2}+1;\frac{1}{2},\frac{3}{2};\tau \right)\\ \nonumber
    &-\frac{1}{2} (n+1) \, _2F_2\left(\frac{n}{2}+1,\frac{n}{2}+\frac{3}{2};\frac{3}{2},2;\tau \right),
\end{eqnarray}
then 
\begin{eqnarray}\label{eq:zeta_inverse}
    \mathcal{L}^{-1}_{s_2\to\tau_2}\left\{\widetilde{\zeta}_n(\sqrt{s_2})\right\}&=\frac{n}{2} (n+1) \, _2F_2\left(\frac{n}{2}+1,\frac{n}{2}+\frac{3}{2};\frac{3}{2},2;\tau_2 \right)\\ \nonumber
    &-\frac{n}{\sqrt{\pi \tau_2 }}\, _2F_2\left(\frac{n}{2}+\frac{1}{2},\frac{n}{2}+1;\frac{1}{2},\frac{3}{2};\tau_2 \right)+\delta(\tau_2),
\end{eqnarray}
where 
\begin{equation}
    \, _2F_2(a,b;c,d,z)=\sum _{k=0}^{\infty } \frac{z^k \left(a_1\right)_k \left(a_2\right)_k}{k! \left(b_1\right)_k \left(b_2\right)_k}, 
\end{equation}
with $(x)_n=\Gamma(x+n)/\Gamma(x)$ \cite{bateman1953higher1}.

From equation (\ref{eq:erfc_series_representation}) we also require, $\mathcal{L}^{-1}_{s_2\to\tau_2}\left\{\widetilde{\zeta}_n(\sqrt{s_2})s_2^{-1/2}\right\}$, where we use the following inversion \cite{roberts1966table}:
\begin{equation}
    \mathcal{L}^{-1}_{s\to\tau}\left\{\frac{s^{n-1}}{(s+1)^n}\right\}= {}_1F_1(n,1,-\tau)
\end{equation}
with equation (\ref{eq:laplace_square_root}) to find,
\begin{eqnarray}\label{eq:zeta_root}
    \fl \mathcal{L}^{-1}_{s_2\to\tau_2}\left\{\widetilde{\zeta}_n(\sqrt{s_2})s_2^{-1/2}\right\} = \frac{1}{\sqrt{4 \pi  \tau_2 ^3}}\int_0^{\infty } u e^ {-\frac{u^2}{4 \tau_2 }} \, _1F_1(n;1;-u) \, du\\ \nonumber
    =\frac{1}{\sqrt{\pi \tau_2} }\, _2F_2\left(\frac{n}{2},\frac{n}{2}+\frac{1}{2};\frac{1}{2},\frac{1}{2};\tau_2 \right)-n \, _2F_2\left(\frac{n}{2}+\frac{1}{2},\frac{n}{2}+1;1,\frac{3}{2};\tau_2 \right).
\end{eqnarray}

Putting all of this together we have
\begin{eqnarray}
    \fl \mathdutchcal{F}_r^+(\tau_1,\tau_2)=\frac{1}{\sqrt{\pi \tau_1}}e^{\tau_2}\mathrm{erfc}(\sqrt{\tau_2})+\sum_{n=0}^{\infty}\frac{(-1)^n \tau_1^{n/2}}{\Gamma(\frac{n}{2}+1)} \mathcal{C}_n^+(\tau_2)+e^{\tau_1}\mathrm{erfc}(\sqrt{\tau_1})\delta(\tau_2)
\end{eqnarray}
where 
\begin{eqnarray}\label{eq:c_plus}
    \fl \mathcal{C}_n^+(\tau_2)=(n+2)\Bigg[\frac{(n+1)}{2}\, _2F_2\left(\frac{n}{2}+\frac{3}{2},\frac{n}{2}+2;\frac{3}{2},2;\tau_2 \right)\\ \nonumber
    -\, _2F_2\left(\frac{n}{2}+\frac{3}{2},\frac{n}{2}+2;1,\frac{3}{2};\tau_2 \right)\Bigg]+\frac{1}{\sqrt{\pi \tau_2}}\Bigg[\, _2F_2\left(\frac{n}{2}+1,\frac{n}{2}+\frac{3}{2};\frac{1}{2},\frac{1}{2};\tau_2 \right)\\ \nonumber
    -(n+1)\, _2F_2\left(\frac{n}{2}+1,\frac{n}{2}+\frac{3}{2};\frac{1}{2},\frac{3}{2};\tau_2 \right)\Bigg].
\end{eqnarray}

\subsection{Double Laplace Inversion of $\widetilde{\mathdutchcal{F}}_r^-(s_1,s_2)$}

We write $\widetilde{\mathdutchcal{F}}_r^-(s_1,s_2)$ as
\begin{equation}
    \widetilde{\mathdutchcal{F}}_r^-(s_1,s_2)=\frac{\sqrt{s_1}+1}{\sqrt{s_1} s_2+\sqrt{s_2} \left(\sqrt{s_1}+\sqrt{s_2}\right)}+\frac{\sqrt{s_2}}{\sqrt{s_1} s_2+s_1 \left(s_2+\sqrt{s_2}\right)},
\end{equation}
then using equations (\ref{eq:laplace_inv_erfc1}) and (\ref{eq:laplace_inv_erfc2}) with (\ref{eq:laplace_deriv}), i.e.
\begin{equation}
    \mathcal{L}^{-1}_{s_1\to\tau_1}\left\{\frac{\sqrt{s_1}}{{\sqrt{s_1}+a}}\right\}=-\frac{a}{\sqrt{\pi \tau_1}}+a^2e^{a^2\tau_1}\mathrm{erfc}\left(a\sqrt{\tau_1}\right) + \delta(\tau_1),    
\end{equation}
we obtain,
\begin{eqnarray}\label{eq:f_minus_mixed}
    \fl \mathcal{L}^{-1}_{s_1\to\tau_1}\left\{\widetilde{\mathdutchcal{F}}_r^-(s_1,s_2)\right\}&=\frac{1}{\sqrt{\pi } \left(\sqrt{s_2}+1\right)^2 \sqrt{s_2 \tau_1}}+\left(1-\frac{1}{\left(\sqrt{s_2}+1\right)^2}\right)\\ \nonumber 
    &\times \frac{e^{\frac{s_2 \tau_1}{\left(\sqrt{s_2}+1\right)^2}} \mathrm{erfc}\left(\frac{\sqrt{s_2 \tau_1}}{\sqrt{s_2}+1}\right)}{\left(\sqrt{s_2}+1\right)}
    +\frac{1}{\sqrt{s_2}+s_2}\delta(\tau_1).
\end{eqnarray}
The Laplace inversion of the first term on the RHS of equation (\ref{eq:f_minus_mixed}) corresponds to equation (\ref{eq:f_minus}). One can see the second term in equation (\ref{eq:f_minus_mixed}) is only different by a factor of $(1+\sqrt{s_2})^{-1}$ to the second term on the RHS in equation (\ref{eq:f_plus_mixed}), then by using equation (\ref{eq:erfc_taylor}), we have
\begin{eqnarray}\label{eq:erfc_series_representation_2}
    \fl \left(1-\frac{1}{\left(\sqrt{s_2}+1\right)^2}\right)\frac{e^{\frac{s_2 \tau_1}{\left(\sqrt{s_2}+1\right)^2}} \mathrm{erfc}\left(\frac{\sqrt{s_2 \tau_1}}{\sqrt{s_2}+1}\right)}{\left(\sqrt{s_2}+1\right)}&=\left(\frac{1}{\sqrt{s_2}}+\frac{\widetilde{\zeta}_1(s_2)}{s_2}\right) \\ \nonumber
    &\times \sum_{n=0}^{\infty} \frac{(-1)^n\tau_1^{n/2}}{\Gamma\left(\frac{n}{2}+1\right)} \widetilde{\zeta}_{n+2}(\sqrt{s_2}).
\end{eqnarray}
While the inversion $\mathcal{L}^{-1}_{s_2\to\tau_2}\left\{\widetilde{\zeta}_n(\sqrt{s_2})s_2^{-1/2}\right\}$ is given by equation (\ref{eq:zeta_root}), we find $\mathcal{L}^{-1}_{s_2\to\tau_2}\left\{\widetilde{\zeta}_n(\sqrt{s_2})/s_2\right\}$ by integrating equation (\ref{eq:zeta_inverse}) over $\tau_2$ to find,
\begin{eqnarray}\label{eq:zeta_integral}
     \mathcal{L}^{-1}_{s_2\to \tau_2}\left\{\frac{\widetilde{\zeta}_n(\sqrt{s_2})}{s_2}\right\} &= \, _2F_2\left(\frac{n}{2},\frac{n}{2}+\frac{1}{2};\frac{1}{2},1;\tau _2\right)\\\nonumber
    &-\frac{2 n \sqrt{\tau _2}}{\sqrt{\pi }} \, _2F_2\left(\frac{n}{2}+\frac{1}{2},\frac{n}{2}+1;\frac{3}{2},\frac{3}{2};\tau _2\right).
\end{eqnarray}
By combining all the above we find $\mathdutchcal{F}_r^-(\tau_1,\tau_2)$ to be,
\begin{equation}
    \fl \mathdutchcal{F}_r^-(\tau_1,\tau_2)=\frac{2}{\pi}\sqrt{\frac{\tau_2}{\tau_1}}-\frac{2\tau_2}{\sqrt{\pi \tau_1}}e^{\tau_2}\mathrm{erfc}(\sqrt{\tau_2})+\sum_{n=0}^{\infty}\frac{(-1)^n \tau_1^{n/2}}{\Gamma(\frac{n}{2}+1)} \mathcal{C}_n^-(\tau_2)+e^{\tau_2}\mathrm{erfc}(\sqrt{\tau_2})\delta(\tau_1),
\end{equation}
where 
\begin{eqnarray}\label{eq:c_minus}
    \fl \mathcal{C}_n^-(\tau_2)=\, _2F_2\left(\frac{n}{2}+\frac{3}{2},\frac{n}{2}+2;\frac{1}{2},1;\tau_2\right)+\frac{1}{\sqrt{\pi \tau_2}}\, _2F_2\left(\frac{n}{2}+1,\frac{n}{2}+\frac{3}{2};\frac{1}{2},\frac{1}{2};\tau_2\right)\\ \nonumber
    \fl -\frac{2(n+3)\sqrt{\tau_2}}{\sqrt{\pi}}\, _2F_2\left(\frac{n}{2}+2,\frac{n}{2}+\frac{5}{2};\frac{3}{2},\frac{3}{2};\tau_2\right)-(n+2)\, _2F_2\left(\frac{n}{2}+\frac{3}{2},\frac{n}{2}+2;1,\frac{3}{2};\tau_2\right).
\end{eqnarray}

\section{Asymptotics of $\mathdutchcal{g}^\pm(\tau_1,\tau_2)$ for $\tau_1,\tau_2\to0$}\label{sec:g_asymptotics}

Using $\lim_{z\to0}{}_2F_2(a,b;c,d;z)=1$ and the definition of $\mathcal{C}^+(\tau_1,\tau_2)$ and $\mathcal{C}^-(\tau_1,\tau_2)$ in equations (\ref{eq:c_plus}) and (\ref{eq:c_minus}) for $\tau_1,\tau_2\to 0$ we find that, $\mathdutchcal{g}^\pm(\tau_1,\tau_2)$ becomes,
\begin{equation}\label{eq:g_plus_sum}
    \mathdutchcal{g}^+(\tau_1,\tau_2)\simeq \sum_{n=1}^{\infty} \frac{(-1)^{n+1}n \tau_1^{n/2}}{\Gamma\left(1+\frac{n}{2}\right)\sqrt{\pi \tau_2}}
\end{equation}
and
\begin{equation}\label{eq:g_minus_sum}
    \mathdutchcal{g}^-(\tau_1,\tau_2)\simeq \sum_{n=0}^{\infty} \frac{(-1)^{n}\tau_1^{n/2}}{\Gamma\left(1+\frac{n}{2}\right)\sqrt{\pi \tau_2}},
\end{equation}
and for $\tau_1\to 0$ the sums in equations (\ref{eq:g_plus_sum}) and (\ref{eq:g_minus_sum}) being dominated by the first term, reduce to
\begin{equation}
    \mathdutchcal{g}^+(\tau_1,\tau_2)\simeq \frac{2}{\pi}\sqrt{\frac{\tau_1}{\tau_2}},
\end{equation}
and 
\begin{equation}
    \mathdutchcal{g}^-(\tau_1,\tau_2)\simeq \frac{1}{\sqrt{\pi \tau_2}}.
\end{equation}

\bibliographystyle{unsrt}
\bibliography{bibliography}
\end{document}